\documentclass{ar-1col}
\usepackage{natbib,todonotes,amsmath,amssymb,caption,subcaption}

\jname{Xxxx. Xxx. Xxx. Xxx.}
\jvol{AA}
\jyear{YYYY}
\doi{10.1146/((please add article doi))}

\begin{document}

\newcommand\notedo[1]{\todo[color=yellow, inline, size=\small]{To do:
    #1}}
\newcommand\notesec[2]{\todo[color=orange, inline, size=\small]{Planned section length: #1 pages.  Status: #2.}}
\newcommand\notesubsec[1]{\todo[color=cyan, inline, size=\small]{Subsection status: #1.}}
\newcommand\maybefig[1]{\todo[color=cyan, inline, size=\small]{Possible figure: #1}}
\newcommand\putcite{\textbf{(CITE!)}}
\newcommand{\beq}{\begin{equation}}
\newcommand{\eeq}{\end{equation}}
\newcommand{\wri}{to write}
\newcommand{\dra}{drafted}
\newcommand{\dracl}{drafted, needs cleanup}

\newcommand{\aj}{AJ}                   
\newcommand{\araa}{ARA\&A}             
\newcommand{\apj}{ApJ}                 
\newcommand{\apjl}{ApJ}                
\newcommand{\apjs}{ApJS}               
\newcommand{\apss}{Ap\&SS}             
\newcommand{\aap}{A\&A}                
\newcommand{\jcap}{J. Cosmology Astropart. Phys.}
\newcommand{\mnras}{MNRAS}             
\newcommand{\nar}{New A Rev.}          
\newcommand{\prd}{Phys.~Rev.~D}        
\newcommand{\prl}{Phys.~Rev.~Lett.}    
\newcommand{\pasp}{PASP}               
\newcommand{\pasj}{PASJ}               
\newcommand{\ssr}{Space~Sci.~Rev.}     
\newcommand{\nat}{Nature}              
\newcommand{\aplett}{Astrophys.~Lett.} 
\newcommand{\physrep}{Phys.~Rep.}   
\newcommand{\procspie}{Proc.~SPIE}   

\markboth{Mandelbaum}{Weak lensing}

\title{Weak lensing for precision cosmology}

\author{Rachel Mandelbaum$^1$
\affil{$^1$McWilliams Center for Cosmology, Department of Physics, 
Carnegie Mellon University, Pittsburgh, PA 15213, USA; email: rmandelb@andrew.cmu.edu}}

\begin{abstract}
  Weak gravitational lensing, the deflection of light by mass, is one of the best tools to constrain
  the growth of cosmic structure with time and reveal the nature of dark energy.  I
  discuss the sources of systematic uncertainty in weak lensing measurements and their theoretical
  interpretation, including our current understanding and other options for future improvement.
  These include long-standing concerns such as the estimation of coherent shears from galaxy images
  or redshift distributions of galaxies selected based on photometric redshifts, along with
  systematic uncertainties that have received less attention to date because they are subdominant
  contributors to the error budget in current surveys.  I also discuss methods for automated
  systematics detection using survey data of the 2020s.  The goal of this review is to describe
  the current state of the field and what must be done so that if weak lensing measurements lead
  toward surprising conclusions about key questions such as the nature of dark energy, those
  conclusions will be credible.
\end{abstract}

\maketitle

\tableofcontents

\section{INTRODUCTION}\label{sec:intro}

\begin{figure}
\centering
\begin{subfigure}{0.5\textwidth}
\centering
\includegraphics[width=0.9\linewidth]{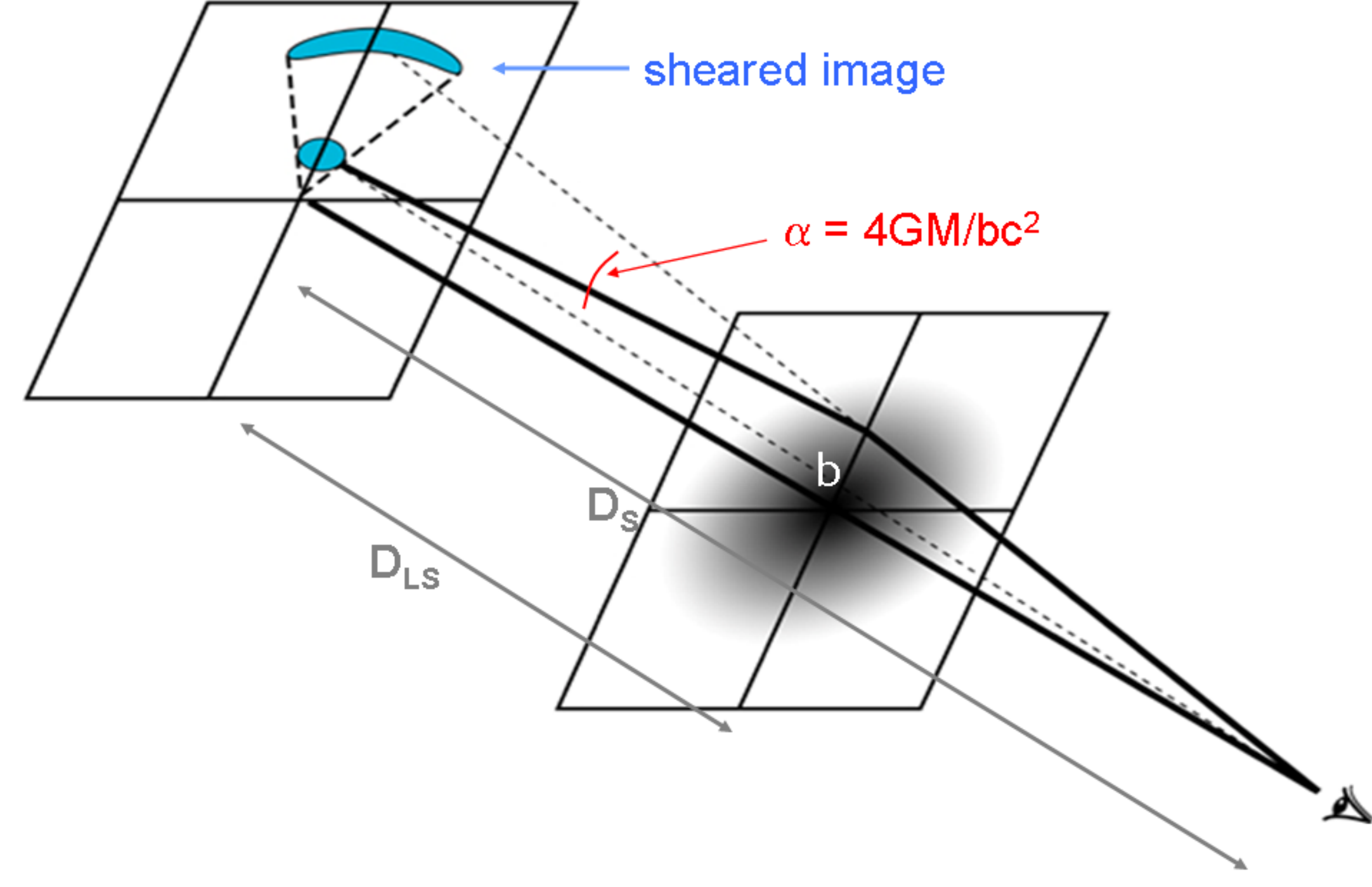}
\end{subfigure}%
\begin{subfigure}{0.5\textwidth}
\centering
\includegraphics[width=0.9\linewidth]{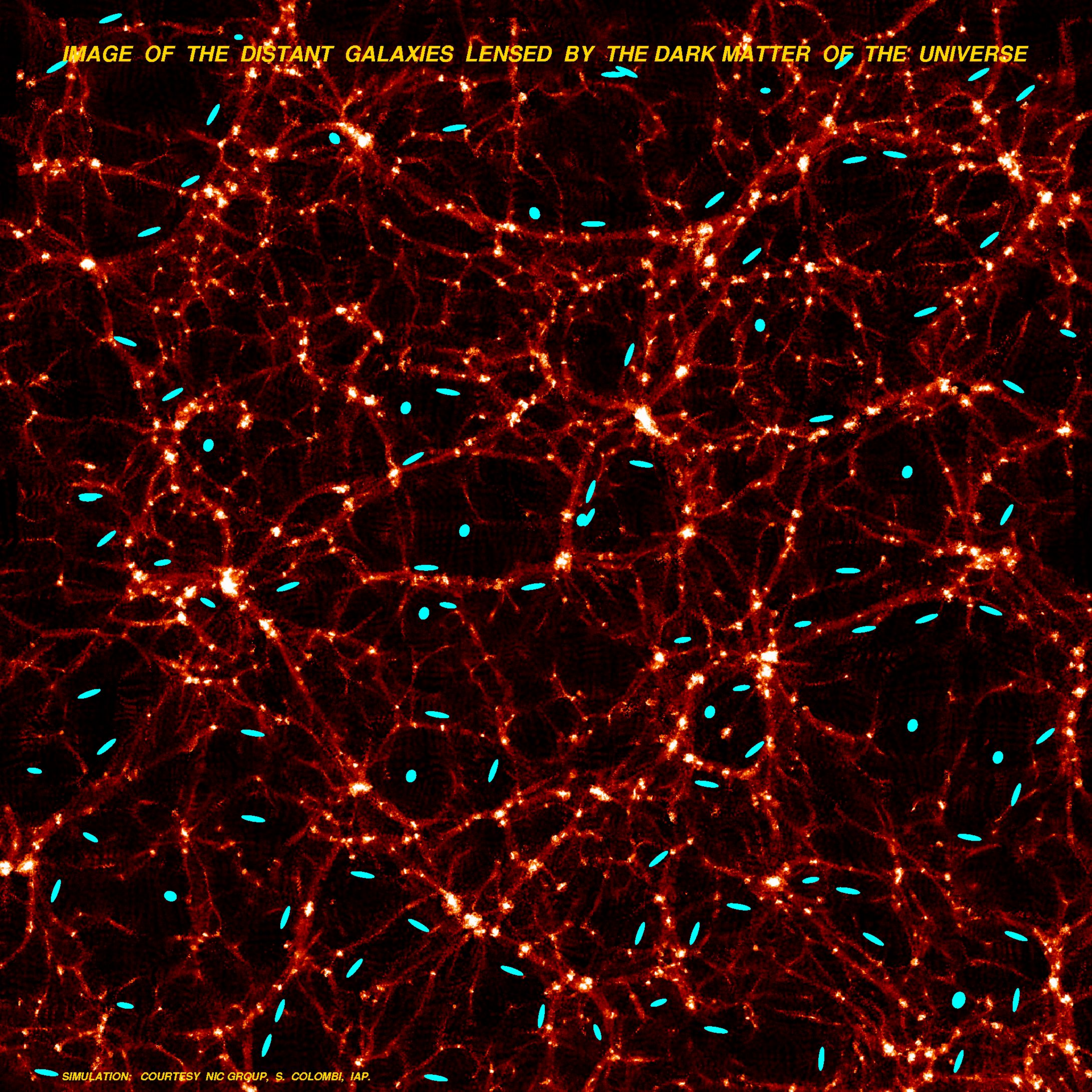}
\end{subfigure}%
\caption{The left panel (from \citealt{2009arXiv0912.0201L}, used with permission from image
  creator, Tony Tyson) illustrates the basic lensing shear
  distortion caused by weak gravitational lensing, for a single lens and source.  The right
  panel (image credit: Canada-France Hawaii Telescope) shows the coherent patterns induced in source
  shapes (blue ellipses) due to large-scale
  structure; the color scale indicates the density in the simulation.}
\label{fig1}
\end{figure}

Gravitational lensing is the deflection of light rays from distant objects by the matter --
including dark matter -- along their
path to us. In the limit that the deflections cause
small modifications of the object properties (position, size, brightness, and shape) but not visually striking
phenomena such as multiple images or arcs, lensing is referred to as ``weak lensing'' \citep[for
recent reviews,
see][]{2015RPPh...78h6901K,dodelsonlensing}.
Since light from distant sources that are near each other on the sky must pass by nearby structures
in the cosmic web (see Figure~\ref{fig1}), their shapes are correlated by lensing.  This correlation
drops with separation on the sky; its amplitude and scale-dependence can be used to infer the
underlying statistical distribution of matter and hence the growth of cosmic structure with time.
This in turn allows us to infer the properties of dark energy \citep[for early work along these
lines, see][]{2002PhRvD..65b3003H, 2002PhRvD..65f3001H}, because the accelerated expansion of the Universe that
it causes suppresses the clustering of matter driven by gravity.

The recognition that weak lensing provides the power to constrain 
the cause of the accelerated expansion rate of the Universe
 has driven the development of ever-larger weak lensing surveys; see e.g., the Dark Energy
Task Force report \citep{2006astro.ph..9591A} and 2010 decadal survey \citep{decadalrev}.  However, weak
lensing can also be used to study the galaxy-dark matter halo connection \citep[e.g.,][]{2015MNRAS.449.1352C,2015MNRAS.447..298H,2015A&A...579A..26V,2016MNRAS.457.3200M} and
constrain neutrino masses \citep[e.g.,][]{2003PhRvL..91d1301A,2011APh....35..177A,2017arXiv170801530D}.

The correlation function of galaxy shapes (shear-shear correlations) is often
referred to as `cosmic shear'. Making these measurements requires (1) the analysis of images to
infer the weak lensing distortions (`shear'), where deviations from purely random galaxy orientations are assumed
to arise due to lensing; (2) estimates of distances to the galaxies involved, in order to
interpret the shape distortions in terms of cosmological parameters; and (3) a host of supporting data e.g.\
to confirm the calibration of  the redshift estimates and inferred shear.
Interpreting them requires the ability to make predictions about the growth of cosmic structure at
late times, well into the nonlinear regime.  

Weak lensing can be described as a linear
transformation between unlensed ($x_u$, $y_u$) and 
lensed coordinates ($x_l$, $y_l$), where the origins of the coordinate systems are at the
unlensed and lensed positions of the galaxy:
\begin{equation}\label{eq:lensingshear}
  \left( \begin{array}{c} x_u \\ y_u \end{array} \right)
  = \left( \begin{array}{cc} 1 - \gamma_1 - \kappa & - \gamma_2 \\ -\gamma_2 & 1+\gamma_1 - \kappa \end{array} \right)
  \left( \begin{array}{c} x_l \\ y_l \end{array} \right).
\end{equation}
There are two components of the complex-valued lensing shear
$\gamma = \gamma_1 + {\rm i}\gamma_2$, which describes the stretching of galaxy images due to
lensing, and the convergence $\kappa$, which describes a change in size and brightness of lensed
objects.    The shear has elliptical symmetry, and hence transforms like a spin-2
quantity.  Since we do not know the unlensed distribution of galaxy sizes 
very precisely, it is common to write this as
\begin{equation}
\left( \begin{array}{c} x_u \\ y_u \end{array} \right)
= (1 - \kappa) \left( \begin{array}{cc} 1 - g_1 & - g_2 \\ -g_2 & 1+g_1\end{array} \right)
\left( \begin{array}{c} x_l \\ y_l \end{array} \right),
\end{equation}
in terms of the reduced shear,
$g_i = \gamma_i / (1 - \kappa)$.

Since the lensing shear causes a change the observed galaxy ellipticities, 
inference of the shear typically depends on measurements of the second moments of galaxies:
\begin{equation}\label{eq:qij}
Q_{ij} = \frac{\int {\rm d}^2 x \, I({\bf x}) W({\bf x}) x_i x_j }
{\int {\rm d}^2 x \, I({\bf x}) W({\bf x}) },
\end{equation}
where $x_1$ and $x_2$ correspond to the $x$ and $y$
directions, $I({\bf x})$ denotes the galaxy image light
profile, and $W({\bf x})$ is a 
weighting function.  One common definition of ellipticity relates to the moments as
\begin{equation}\label{eq:ellipticity}
e = e_1 + {\rm i} e_2 = \frac{Q_{11} - Q_{22} + 2 {\rm i} Q_{12}}{Q_{11} + Q_{22}}.
\end{equation}
Another definition of ellipticity replaces the denominator in
Eq.~\ref{eq:ellipticity} with $Q_{11} + Q_{22} + 2 [\text{det}(Q)]^{1/2}$.  Both ellipticity definitions 
have a well-defined response to a lensing shear, and hence can be averaged across
ensembles of galaxies.  A variety of
methods exist for estimating these ellipticities while removing the effect of the point-spread
function (PSF) from the atmosphere and telescope.

The convergence can be thought of as the projected matter overdensity, defined for a given point on
the sky $\mathbf{\theta}$ as
\begin{equation}
\kappa(\mathbf{\theta}) = \frac{3H_0^2 \Omega_m}{2c^2} \int_0^\chi \frac{\mathrm{d}\chi \,\chi
  \, q(\chi)}{a(\chi)} \delta(\chi\mathbf{\theta}, \chi)
\end{equation}
in a flat Universe.  Here $H_0$ is the current Hubble parameter, $\Omega_m$ is the current matter
density in units of the critical density, $a$ is the scale factor, $\delta$ is the matter
overdensity, $\chi$ is the comoving distance, and the lens efficiency $q$ is defined as
\begin{equation}
q(\chi) = \int_\chi^\infty \mathrm{d}\chi' n(\chi') \frac{\chi'-\chi}{\chi'}
\end{equation}
in terms of the source distribution $n(\chi')$.  The line-of-sight projection in the
expression for $\kappa$ indicates why the most interesting weak lensing measurements involve binning
by source redshift (``tomography''): instead of averaging over all line-of-sight structure, using a
set of distinct redshift bins enables measurement of how cosmic structure has grown with time.

From the  initial detections of
  cosmic shear \citep{2000MNRAS.318..625B,
    2000A&A...358...30V,2000Natur.405..143W,2001ApJ...552L..85R} to recent measurements
  \citep[e.g.,][]{2016PhRvD..94b2002B,2016ApJ...824...77J,2017MNRAS.465.1454H,2017arXiv170801538T},
  there has been substantial evolution of methodology to ensure that either observational or
  astrophysical systematic
  errors do not dominate the measurements.  Currently, three weak lensing surveys are
  ongoing:  the
  Kilo-Degree Survey (KiDS; \citealt{2013ExA....35...25D}),
  the Dark Energy
  Survey (DES; \citealt{2016MNRAS.460.1270D}), and the Hyper Suprime-Cam
  survey (HSC; \citealt{SurveyOverview}).  In the 2020s, several ``Stage-IV'' \citep{2006astro.ph..9591A}
  surveys will further increase the precision of these measurements:  Euclid
  \citep{2011arXiv1110.3193L}, LSST 
  \citep{2009arXiv0912.0201L}, and WFIRST 
  \citep{2015arXiv150303757S}.

\begin{figure}
\centering
\begin{subfigure}{0.5\textwidth}
\centering
\includegraphics[width=0.9\linewidth]{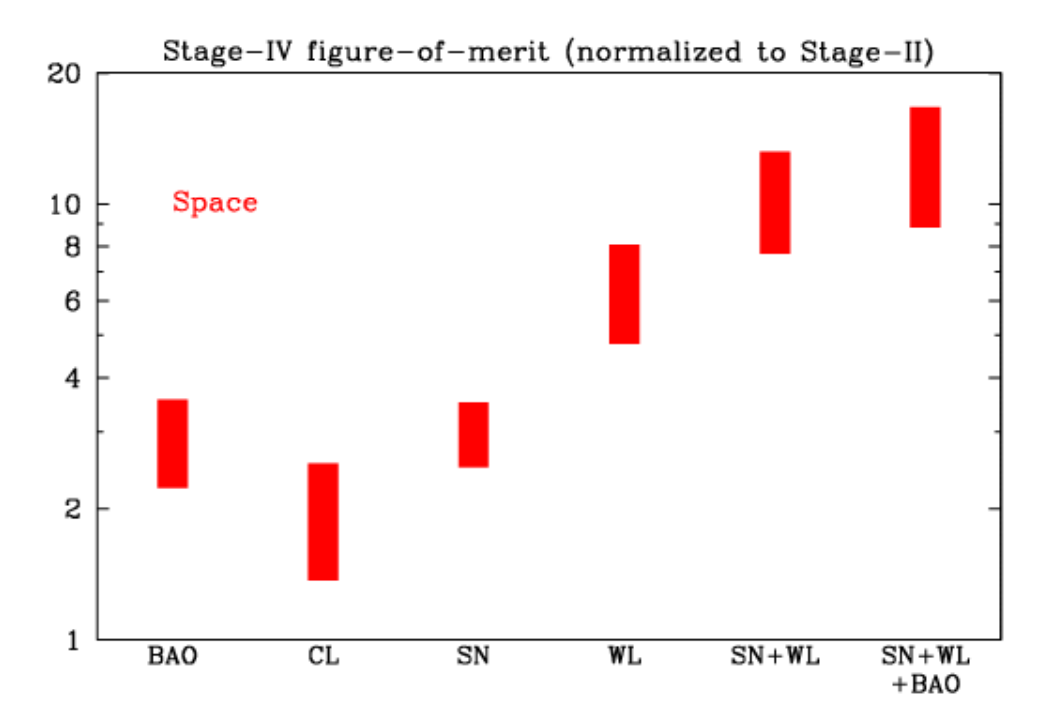}
\end{subfigure}%
\begin{subfigure}{0.5\textwidth}
\centering
\includegraphics[width=0.9\linewidth]{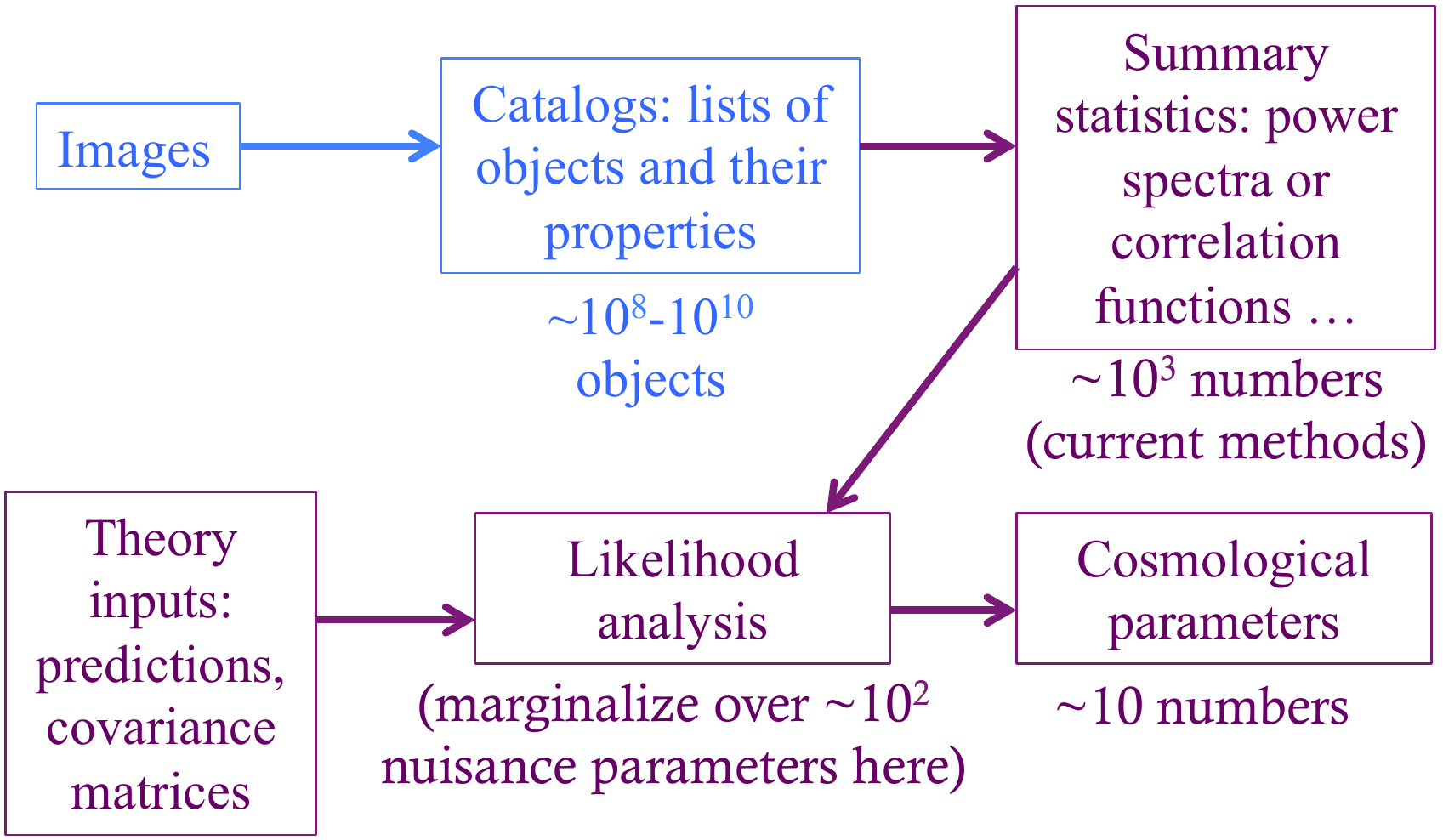}
\end{subfigure}%
\caption{\textbf{Left:} Forecast of dark energy constraining power in Stage-IV space-based 
surveys of the 2020s,
  normalized with respect to the Stage-II surveys that existed in $\sim$2005 (from
  \citealt{2006astro.ph..9591A}). The increase in this
  figure-of-merit is based on the ability to constrain the equation of state of dark energy,
  including its time evolution, for baryon acoustic oscillations (BAO), galaxy clusters (CL), Type Ia
  supernovae (SN), weak lensing (WL), and two probe combinations. While the exact details for actual
  Stage IV surveys will depend on the survey design, this figure nonetheless demonstrates the basic
  principle that has driven much of the excitement about weak lensing in the cosmological
  community. \textbf{Right:} Basic outline of the weak lensing analysis process, where the blue
  and dark purple respectively indicate the parts of the analysis covered in Sections~\ref{sec:im2cat}
  and~\ref{sec:cat2science} of this review.}
\label{fig:forecast}
\end{figure}

  The focus of this review is on weak lensing method development and systematics mitigation in
  preparation for the surveys that will happen in the 2020s, which will have such small statistical
  errors that serious discrimination among dark energy models will be possible, the era of
  ``precision cosmology'' (Figure~\ref{fig:forecast}, left panel).  Since the weak lensing shear is so small compared to the intrinsic,
  randomly-oriented 
  galaxy ellipticities (often called ``shape noise''), averaging over very large ensembles of
  galaxies is the key to achieving small statistical errors.  Indeed, this shape noise dominates
  over the impact of pixel noise in galaxy shape estimation for nearly all galaxies above
  detection significance of $\sim$5.  Hence weak lensing measurements generally use galaxies that
  are as faint and small as possible, down to the limit imposed by the need to control systematic
  uncertainties.  I describe the obstacles in the path towards a statistical error-dominated
  analysis, status of existing analysis methods, and areas where more work is needed.  How do we do
  weak lensing correctly when we need to trust it for potentially novel results, such as unusual
  findings about dark energy?  The goal of the review is to explain the technical state of the art
  and challenges moving forward towards this goal.

This review will cover both observational and theoretical systematics in weak
lensing two-point correlations, both shear-shear (cosmic shear) and shear-galaxy (``galaxy-galaxy
lensing'').  The canonical cosmological weak lensing analysis from Stage IV surveys will include
joint analysis of shear-shear, shear-galaxy, and galaxy-galaxy correlations; however, I 
refer the reader to other works for thorough discussion of systematics in galaxy-galaxy correlations
\citep[e.g.,][]{2015MNRAS.454.3121M}.  Throughout this work, I refer to two-point correlations generically; in practice, they
may be estimated in configuration space (``correlation functions'') or Fourier space (``power
spectra'').  The review will cover the entire path from raw images to science; see analysis
flowchart in the right panel of Figure~\ref{fig:forecast}.  It will
focus on cosmological distance scales, and will not cover the possibility of using small-scale
lensing and clustering \citep[e.g.,][]{2015ApJ...806....2M}, which brings in a host of additional theoretical and observational
issues, or cluster number counts \citep[e.g.,][]{2015MNRAS.449..685H}.

To enable thorough discussion of the above topics, several other 
approaches to weak lensing analysis will be neglected.
These include shear beyond two-point correlations; flexion; lensing magnification;
lensing cosmography (constraints on distance ratios
rather than structure growth); and weak lensing outside of
the optical or NIR wavelength range, such as radio lensing.
Lensing of the Cosmic Microwave Background
\citep[CMB; e.g.,][]{2014A&A...571A..17P} will
be discussed as a consistency check on optical lensing, without coverage of 
its systematic uncertainties. 


The structure of this review is as follows.  I divide the weak lensing analysis process into two
major steps: from images to catalogs with measured object properties (Section~\ref{sec:im2cat}), and from
catalogs to cosmological parameters (Section~\ref{sec:cat2science}).  
The additional step of detecting and controlling for observational systematics more generally
is described in Section~\ref{sec:obssys}.  I summarize the future prospects for the field in Section~\ref{sec:summary}.


\section{FROM IMAGES TO CATALOGS}\label{sec:im2cat}


The full weak lensing analysis process goes all the way from the raw pixel data to cosmological
parameter constraints.  While somewhat artificial, it is common to separately consider the analysis
steps from images to catalogs, followed by catalogs to science.  This division partly reflects where
certain systematics can be mitigated.  Systematics related to the measurement process 
have a first-pass correction during the ``images to catalogs'' pipeline,
and any mitigation for insufficiency of those corrections occurs in the ``catalogs to science''
pipeline.  Theoretical systematics, i.e., those related to our insufficient knowledge of
astrophysics, can only be mitigated in that second step.  Hence we can think of the ``images to
catalogs'' pipeline as the place where the weak lensing community attempts to estimate the
properties of astronomical objects as accurately as possible, and the ``catalogs to
science'' pipeline as the place where all residual systematics must get mitigated.

\begin{figure}
\begin{center}
\includegraphics[width=\linewidth]{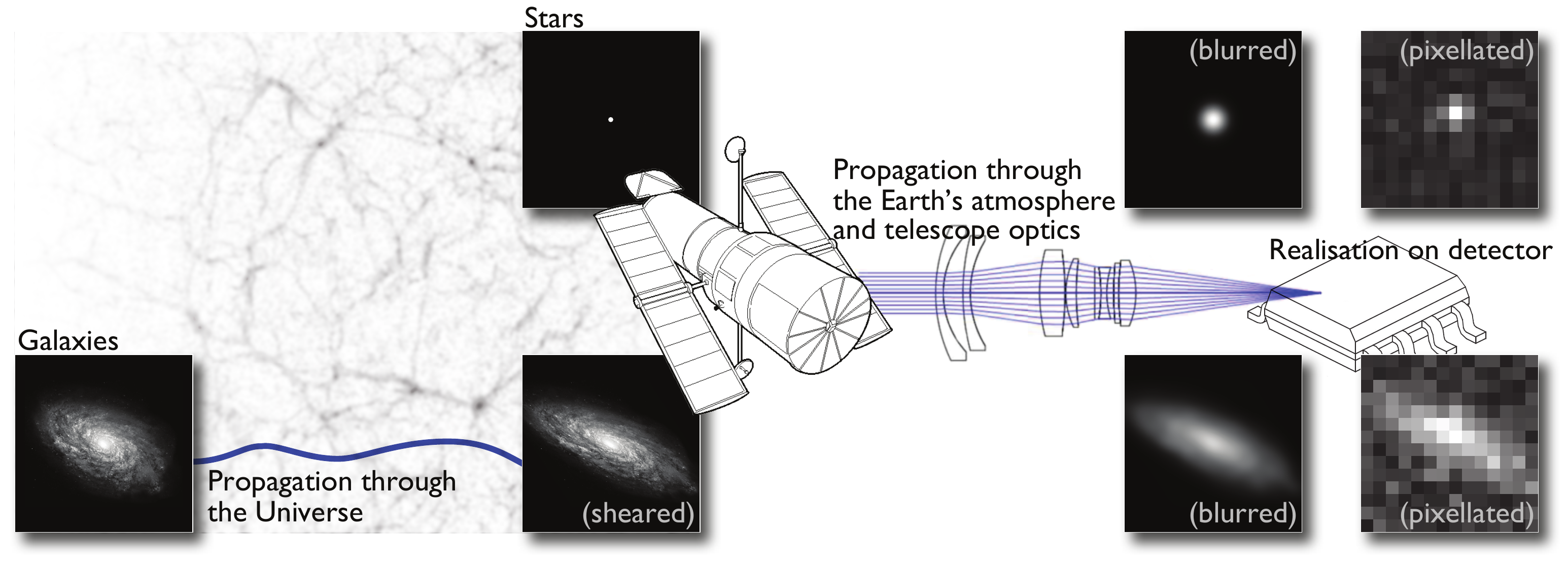}
\end{center}
\caption{An illustration of the processes that affect the galaxy image (from
  \citealt{2014ApJS..212....5M}), including lensing followed by other effects that also
  cause coherent shape distortions, such as convolution with the point-spread function (or PSF) due
  to the atmosphere (for ground-based telescopes) and telescope optics.}
\label{fig2}
\end{figure}

This 
section focuses on the first part of that pipeline.  The basic challenge of inferring the
lensing shear from galaxy images is illustrated in Figure~\ref{fig2}.  I will describe the nature of the
challenges that arise at each step, the state of the art, and directions for future work.


\subsection{PSF modeling}\label{subsec:psfmodeling}


In this review, I consider the atmospheric PSF, optical PSF, pixel
response, and charge diffusion together as the effective PSF.  In other words, the model is that the pixel response
(ideally a top-hat function, though reality can be more complex, resulting in higher-order corrections) is
convolved with the other PSF components, and then the image is sampled at pixel
centers\footnote{This model is violated if the pixels do not form a regular grid; current
  estimates suggest that the impact of deviations from a regular grid in real sensors are sufficiently
  small to ignore even for LSST \citep{2017PASP..129h4502B}}.

Inferring coherent weak lensing distortions requires correction for the effect of the PSF, which if
insufficiently corrected (a) dilutes the shear estimates, causing a multiplicative bias that is worse for small
galaxies, and (b) imprints coherent additive corrections 
to the galaxy ellipticity values, due to the PSF anisotropies.  Methods for removing the impact of the PSF
from shear estimates all come with the assumption that the PSF is known.  Hence, modeling the
PSF correctly is an important challenge for weak lensing; errors in PSF model size and shape result
in multiplicative and additive shear biases, respectively.  The exact impact on the
ensemble weak lensing shear observables depends not just on mean PSF size or shape errors, but
rather their spatial correlations, and the distribution of galaxy properties.  The formalism for understanding these effects
either through simulations or through a moments-based formalism for propagating PSF
modeling errors into shear biases was developed over the past decade 
\citep{2004MNRAS.347.1337H,2004MNRAS.353..529H,2006JCAP...02..001J,2008A&A...484...67P,2010MNRAS.404..350R}.  While
accurate determination of the PSF model is critical for weak lensing cosmology, it
is fortunately a systematic uncertainty that comes with null tests that can be
used to empirically identify problems, drive
algorithmic development, and derive bias corrections.  

\begin{figure}
\begin{center}
\includegraphics[width=0.75\linewidth]{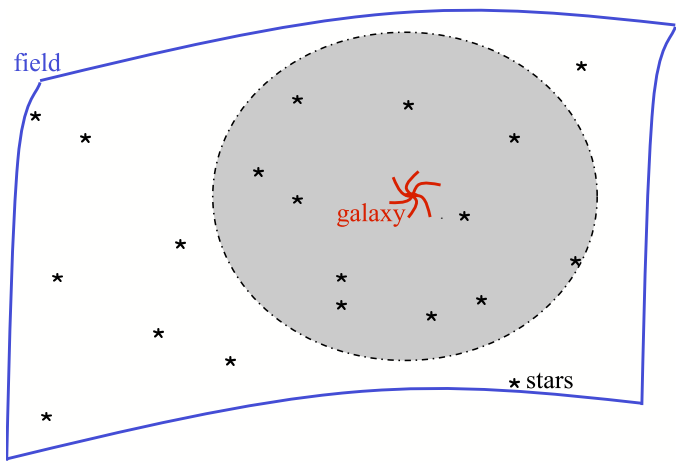}
\end{center}
\caption{An illustration of the PSF interpolation problem 
  (credit: \citealt{2008A&A...484...67P}, A\&A, 484, 67, reproduced with permission \textcopyright{} ESO).  From a sparse sampling of stars in a given field, the PSF
  must be interpolated to the positions of galaxies so their properties can be measured.}
\label{fig3}
\end{figure}
\begin{figure}
\begin{center}
\includegraphics[width=0.75\linewidth]{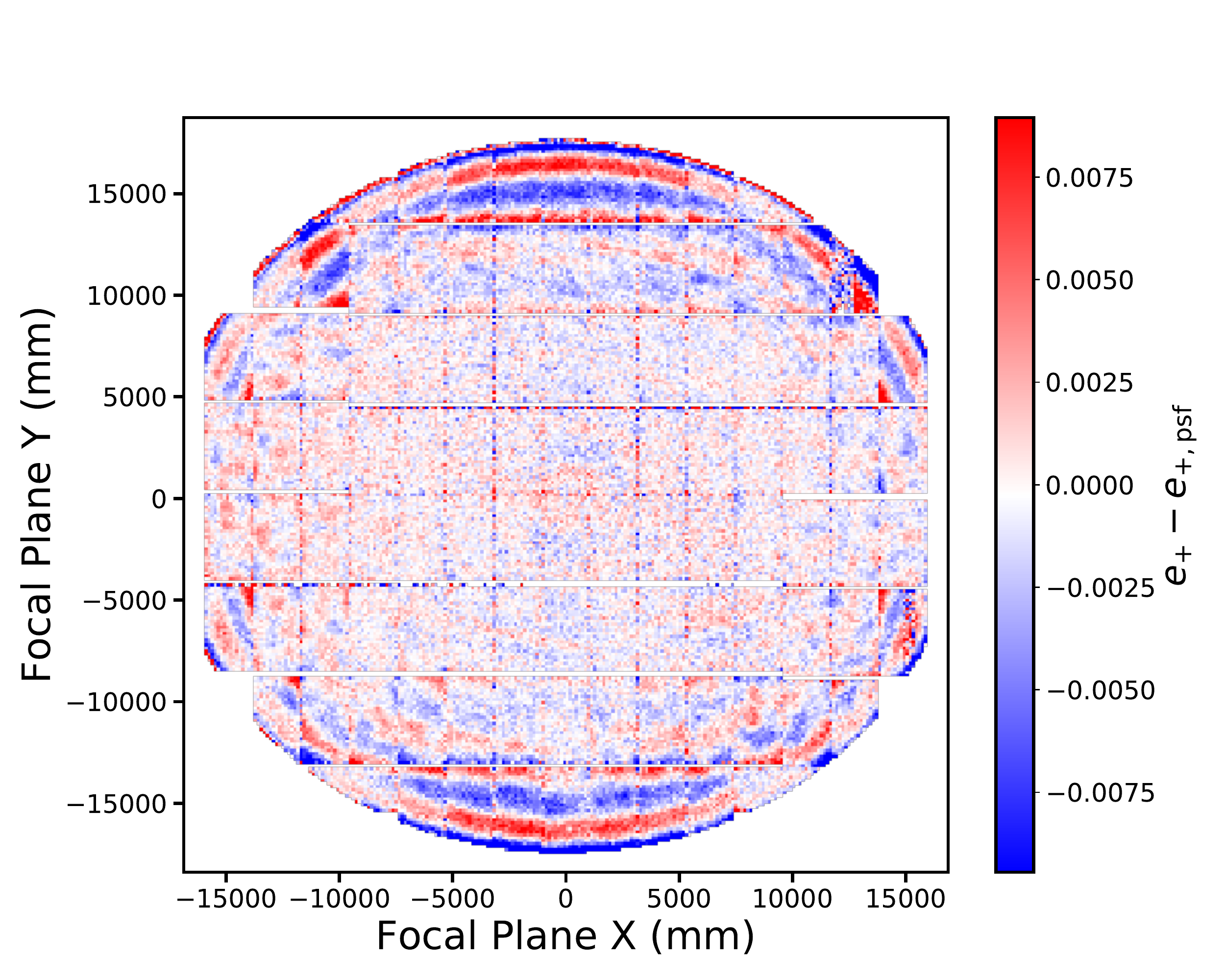}
\end{center}
\caption{The color scale shows
  the PSF model radial ellipticity residual ($\Delta e_+$) averaged over many 
  HSC survey exposures.  Here `radial' refers to the ellipticity component defined with respect to
  the focal plane center.  The rings of nonzero values indicate a coherent 
misestimation of the radial ellipticity of the PSF near the focal plane edge
edge. Figure provided by Bob Armstrong, based on figures and data from \citealt{PipelinePaper:inprep}. }
\label{fig4}
\end{figure}

In principle, the PSF modeling process may be thought of as having two components (see
Figure~\ref{fig3}).  The first is using bright star images to model the PSF. 
The second is
interpolating to other positions so the PSF model can be used to measure galaxy photometry and
shapes \citep[for discussion of several PSF interpolation methods, see
e.g.][]{2012MNRAS.419.2356B,2013A&A...549A...1G,2013ApJS..205...12K,2017AJ....153..197L}.  Challenges in PSF modeling
and interpolation differ for ground- and space-based imaging.  The optical PSF can be thought of as
varying slowly over the field-of-view and exhibiting a limited set of predictable patterns, which is one of the appeals of
space-based weak lensing measurements.  In contrast, the atmospheric PSF
exhibits stochastic behavior (for which the power spectrum can be measured, and each exposure is a
different realization of that power spectrum).  

Some PSF modeling and interpolation methods are purely empirical. 
These involve choosing a set of basis functions to describe the bright star images, and
some functions for interpolating
between those images within a single CCD chip  (e.g., regular or Chebyshev polynomials, though more
sophisticated options exist).  The PSF tends to exhibit discontinuities at chip
boundaries due to slight inconsistencies in chip heights, which makes modeling purely within chips a
common process.  An example of an empirical PSF modeling algorithm is \texttt{PSFEx} \citep{2011ASPC..442..435B}, which
was used for both DES and HSC.  Figure~\ref{fig4} shows a typical failure mode for empirical
approaches: failure to properly describe PSF variations in parts of the focal plane
with the adopted interpolation functions.

One method that has the potential to address both the PSF modeling and interpolation problems is Principal
Component Analysis, or PCA \citep{2007PASP..119.1403J,2010A&A...516A..63S}.  The
PCA method considers all of the survey data, and identifies the most important patterns in
PSF model variation across that data.  PCA analysis can be done at the level of PSF images or 
any compact representation of the PSF, such as its second moments.  Due to its use of all
survey data, with stars in different exposures sampling different locations in the focal plane, the
method can determine PSF model variation as a function of focal plane position at higher spatial
frequency than is possible using only the stars observed on a single exposure.  
Naively, this is a more promising method for space-based data (which only has an optical PSF
determined by a relatively limited set of physical parameters).

Another approach to PSF modeling is a physics-based forward-modeling approach.  One past example of
this (for the {\em Hubble Space Telescope}, or {\em HST})
is ray-tracing through a physical model of the telescope optics using \texttt{Tiny
  Tim}\footnote{\texttt{http://www.stsci.edu/software/tinytim/}}.  This was used by the COSMOS team
\citep[e.g.,][]{2007ApJS..172..219L} for several of its analyses.  The idea is to
forward-simulate the PSF as a function of position in the focal plane in each band given a limited
set of physical parameters such as variation in focus position, then match the stars in each
exposure against those models to identify the best model for each exposure.  The PSF model
interpolation then uses the finely-spaced grid of models rather than the more widely-spaced
stars.   The most obvious
failure mode for forward-modeling of the PSF is if some relevant physics determining the PSF is not
included in the model \citep[for an example of this in practice, see][]{1998SPIE.3355..608S}. 

While physical modeling seems most appropriate for space-based PSF modeling and
interpolation, in principle one approach for LSST is a combination of an optical model
(perhaps with additional empirical constraints from wavefront data,
e.g.~\citealt{2014SPIE.9145E..16R} and \citealt{2016SPIE.9906E..4JX}) and a stochastic atmospheric
PSF model using, for example, Gaussian processes or a maximum entropy algorithm \citep{2012MNRAS.427.2572C}. 
Empirical modeling of the optical PSF using wavefront measurements from out-of-focus
exposures at least partially mitigates the concern about missing physics in the pure forward-modeling
approach.
One important advantage of combined optical plus atmospheric PSF modeling is that the optical component
can potentially include the chip discontinuities, enabling the atmospheric model to use the entire
focal plane rather than modeling each chip separately.

Recently, work on PSF modeling systematics has gone beyond second
moments-based size and shape estimates.  Getting higher order moments of the PSF model
wrong may be problematic; such errors can be identified most easily by comparing stacked star images and PSF
models to identify differences that are not easily described using the second
moments.  Quantifying their impact on weak lensing is most easily done with simulations; no
simple analytic formalism has been worked out in this case.  Also,
failure of the PSF and galaxy profiles to be well-approximated by a Gaussian (e.g., like space-based
PSFs, since the Airy function has a formally infinite variance) causes the simple analytic formalism
for second moments to fail, rendering simulations necessary.

Another effect that has gotten more attention in recent years is the 
chromatic PSF.  Both the atmospheric and optical PSFs depend on wavelength; even the
sensor contribution to the PSF can exhibit slight wavelength dependence as well \citep{2015JInst..10C6004M}.  Within a
single broad photometric band, the effective PSF for any given object must depend 
on its spectral energy distribution (SED).  Since stars and galaxies tend to have different SEDs,
they will have different effective PSFs, which is a problem when using star images to infer the PSF
for galaxies.  Even worse, galaxy color gradients cause a 
violation of the assumption that there is a single well-defined PSF for the galaxy.  Substantial
work has been done on the chromatic PSF effect on weak lensing measurements
\citep{2010MNRAS.405..494C,2012PASP..124.1113A,2012MNRAS.421.1385V,2015ApJ...807..182M,2017arXiv170806085E}.
While the magnitude of the effect tends to be larger for space-based PSFs, for which the relevant physics scales like
$\lambda$ rather than $\sim\lambda^{-1/5}$ for atmospheric PSFs, the actual importance of the effect
for science depends on the requirements on how well PSF size is known.  These requirements may be stricter for ground-based surveys given their
larger PSF size.  Broader bands, such as those planned
for the Euclid survey, are more problematic in this regard.  The above references include work on mitigation schemes that approach the level of systematics control needed
for future surveys.  

As mentioned above, there are
well-defined null tests that can directly reveal PSF modeling errors, unlike
some of the other systematic uncertainties described in this review.  These null tests typically
involve sets of stars (high-significance detections) that were not used for PSF modeling.  A
comparison of their sizes and shapes based on second moments with the PSF model sizes and shapes at
the positions of those stars can be quite revealing.  While the most obvious thing to do is make a
histogram of those differences and look for systematic biases, the spatial
correlation function of these errors determines how weak lensing observables will be biased due to PSF
modeling errors.  For PSF shape errors, there are five relevant correlation functions (called
$\rho$ statistics), two introduced
by \citet{2010MNRAS.404..350R} and three by \citet{2016MNRAS.460.2245J}; these include factors of the PSF shape residuals, the PSF shape
itself, and the PSF size residuals, and directly correspond to additive terms in the shear-shear
correlation function generated by PSF modeling errors.  For examples of their use in real survey data, see
\citet{2016MNRAS.460.2245J,CatalogPaper:inprep}. 

An additional diagnostic is to compare the distribution of PSF shape and size errors for the non-PSF
star sample with the distribution for those stars used to estimate the PSF.  If the two samples have
the same detection significance, then the widths of the distributions can reveal whether there are overall PSF modeling
issues (similar breadth of the distributions) or whether there may be issues with overfitting or
interpolation (broader distribution for non-PSF stars). Comparison of the $\rho$ statistics computed with PSF and
non-PSF stars can also be revealing.
 Finally, stacking the PSF size or shape
residuals in the focal plane coordinate system, across multiple exposures, can reveal systematic
failure to model recurring optical features in the PSF; see Figure~\ref{fig4} for an example.

Aside from the obvious approach of developing more sophisticated PSF modeling algorithms, survey
strategy may mitigate the impact of single-exposure PSF modeling errors on the
final multi-exposure shear estimate.  For example, consider the not atypical case that PSF modeling errors
systematically correlate with distance from the center of the focal plane (e.g., Figure~\ref{fig4}).  If all exposures in
that region have very small dithers, then each galaxy will be observed at nearly the same focal
plane position in all exposures, and
their PSF modeling errors will be coherent. If there is substantial
dithering compared to the size of the focal plane, then the galaxy will be observed at many
different focal plane positions, and the systematics due to PSF modeling errors will
average down.  Depending on the coherent structure of PSF modeling errors, rotational dithering
may also be beneficial.  Using survey strategy to reduce systematics in large-scale
structure statistics was considered by \citet{2016ApJ...829...50A} for LSST, where hundreds of
exposures make this approach to systematics mitigation possible \citep{2017arXiv170804058L}, but a similar study
in the weak lensing context has not yet been performed.

\subsection{Detector systematics}\label{subsec:detector}


For the purpose of weak lensing, detector non-idealities can cause two problematic types of systematics
that cannot be treated as a simple convolution (and hence as part of the 
PSF).  First, there are flux-dependent effects that predominantly affect bright objects, such as nonlinearity or
the brighter-fatter effect discussed below.  Since weak lensing measurements are dominated by faint
galaxies, but the PSF for those faint galaxies is estimated via interpolation between the PSF
modeled from bright stars, detector non-idealities affecting bright objects result in the wrong PSF
being used when estimating shear from the faint galaxies\footnote{Technically, without correction
  for the brighter-fatter effect, the PSF estimated
  from the bright stars is not even the right PSF to use for the bright galaxies.}.  The second type of detector
non-idealities affect all objects, due to defects that correlate with position or galaxy
orientation on the detector.  They can induce spurious coherent shear signals or photometry errors,
and/or cause selection biases due to coherent masking patterns.
While correction for some detector non-idealities such as nonlinear response has long been taken for
granted as happening before the stage that weak lensers care about, the
field's approach to detector systematics has otherwise been varied.

For example, the detectors on the {\em HST} are known to suffer from charge
transfer inefficiency (CTI) due to radiation damage.  CTI imparts a preferential
direction in the images, which is a problem for weak lensing measurements, the goal of which is to
identify coherent smearing in galaxy shapes.
A physically-motivated pixel-level correction scheme 
was pioneered primarily by and for weak lensers, resulting in a 97\% correction for this
effect \citep{2010MNRAS.401..371M,2010PASP..122..439R}.

In the past few years, there have been many more studies on the detailed impact of
detector non-idealities on weak lensing.  One example is the so-called
``brighter-fatter'' effect \citep{2014JInst...9C3048A,2015A&A...575A..41G}, wherein brighter objects
spuriously appear slightly broader than fainter ones due to the electric
field sourced by charges accumulated within a pixel deflecting later light-induced charges away from
that pixel.  
 Conceptually, one can think of this effect as a dynamic change in pixel
boundaries.  While early work proposed methods for estimating the effect using flat
fields, later work has focused on detailed simulation and measurement methods
\citep[e.g.,][]{2017JInst..12C3091L}.  This effect is quite problematic for weak lensing
because the if left uncorrected, the PSF inferred from bright stars is not the relevant one to
use when removing the impact of the PSF on the faint galaxies that dominate weak lensing
measurements.  Fortunately, empirical tests of PSF model fidelity can be carried out
as a function of magnitude to confirm that the brighter-fatter effect has been corrected at the
necessary level. These were used in the HSC survey to identify the impact of the 
brighter-fatter effect, and show that the corrections were sufficient for weak lensing science 
in HSC \citep{CatalogPaper:inprep}.

The conceptual framework mentioned above, wherein some detector non-idealities are thought of as
dynamically adjusting pixel boundaries and therefore pixel sizes (resulting in 
astrometric and photometric errors), applies to several other
detector effects.  For example, the concentric rings known as ``tree rings'' and bright stripes near
detector edges known as ``edge distortions'' in DES can be modeled this way.
\citet{2014PASP..126..750P} proposed that the templates for these
effects derived using flat-field images can be used in the derivation of photometric and
astrometric solutions.  In other words, the WCS (world coordinate system) that maps from image to
world coordinates can include these (admittedly rather complex) effects
\citep{2014SPIE.9150E..17R,2017PASP..129h4502B,2017arXiv170609928B}.  Note that modeling the effect
as part of the WCS is a distinct solution from pixel-level correction, such as was used for CTI; the
WCS-based correction kicks in when measuring positions, photometric quantities, and galaxy shapes
from the images.  It is a valid approach in the limit that the detector effect can be described with a WCS that
varies slowly compare to the size of a pixel, though since it is common practice to take the local
affine approximation of the WCS when measuring individual objects, that imposes a more stringent
constraint that the WCS can be considered locally affine over the scale of individual objects.

The details of detector non-idealities depend on the detectors used for
each survey (though common mitigation schemes can be used for conceptually similar systematics in
multiple surveys).  The general
framework for how detector effects impact weak lensing developed in \citet{2013MNRAS.429..661M} for
Euclid would be relevant for other surveys, with the exact effects to be considered varying. 
A new complication is the fact that WFIRST will use near-infrared
(NIR) detectors, which operate differently from CCDs. CMOS devices have 1 readout path per pixel,
whereas CCDs have 1 readout path per channel; calibrating all pixels to within requirements,
including the effect of cross-talk, is more challenging for CMOS devices. The use of
different types of detectors necessitates 
studies of the impact of various NIR detector effects, some of which are present in CCDs
(e.g., nonlinearity and brighter-fatter: \citealt{2016PASP..128j4001P,2017JInst..12C4009P}), and others that are not,
such as interpixel capacitance \citep[IPC:][]{2016PASP..128i5001K}, persistence, and correlated read noise.  Further work
on characterizing the impact of NIR detector systematics for weak lensing is underway, with an eye
towards placing requirements on hardware and survey strategy to ensure that residual systematics
can be mitigated at the level needed for weak lensing with WFIRST.  

A range of correction schemes have been discussed for various detector effects, including 
pixel-level correction, including them in the WCS, and
applying catalog- or higher-level mitigation schemes
such as template marginalization.  Understanding the spatial- and time-dependence of detector
effects is also quite important, and can be a challenge especially for CMOS detectors.
 In principle there may also be the option of
indirect mitigation through survey strategy for effects that correlate with location on the focal
plane \citep[e.g., following the approach of][]{2016ApJ...829...50A}. 
Additional work is needed to quantify the impact of various low-level detector systematics for
upcoming surveys, including lab measurements and simulations of their impact on weak lensing.
Detectors are sufficiently complex, and requirements on systematics sufficiently strict for upcoming
surveys, that analysis of realistic lab data is a necessity to avoid unpleasant surprises during
commissioning -- with ongoing efforts in both WFIRST
and LSST \citep[e.g.,][]{2013PASP..125.1065S,2014SPIE.9154E..15T}.

\subsection{Detection and deblending}\label{subsec:deblending}


Traditionally, object detection is carried out by detecting peaks above some detection threshold.  For weak lensing, additional cuts are typically placed to identify objects that
can be well-measured; these cuts can be a source of ``selection bias'' (see
Section~\ref{subsec:selbias}). 

\begin{figure}
\begin{center}
\includegraphics[width=0.75\linewidth]{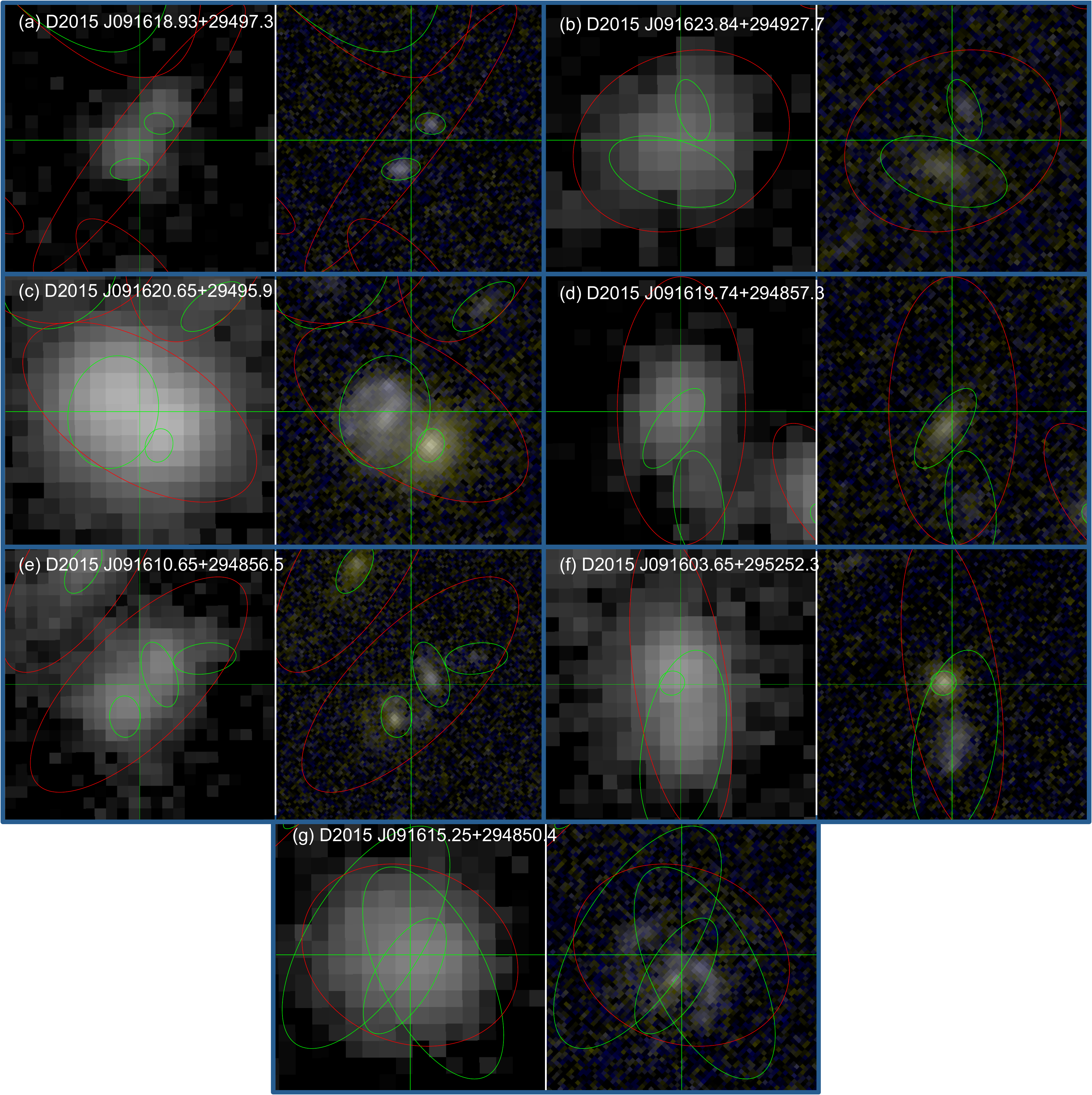}
\end{center}
\caption{An illustration of the issues with unrecognized blends (\textcopyright{} AAS.  Reproduced with permission, from \citealt{2016ApJ...816...11D}).
  Each pair of images shows a ground-based (left) and space-based (right) image of the same system,
  with the shapes of the galaxy detection in the ground-based and space-based images shown as red and green
  ellipses.}
\label{fig5}
\end{figure}

``Deblending'', the process of removing the influence of light from other objects above
that same detection threshold, requires the identification of detections that have multiple peaks.
This naturally leads to two regimes: recognized blending, wherein
the multiple peaks are recognizable, and unrecognized\footnote{Terminology for these varies; e.g.,
  they are called ``ambiguous'' blends
in \citet{2016ApJ...816...11D}.} blending, wherein the deblending algorithm is not
triggered because multiple peaks are not identified within the detection  (see
Figure~\ref{fig5}).  The same system could switch between these categories depending on the
PSF size.  In the case of mild
blending, one can ask whether the deblending algorithm results in unbiased
measurements of object properties, or whether there are coherent systematics requiring mitigation
and/or removal of mildly blended objects.  For unrecognized blends, the only possibility is to quantify their
rate of occurrence, and apply analysis-level mitigation strategies.

For weak lensing, the primary concerns are the impact of blending on shear and photometric
redshifts.  If we consider unrecognized blends, then two objects at the same redshift should have the
same shear, and therefore it should be possible to properly calibrate shear estimates for the combined
(non-deblended) object.  However, for photometric redshifts of unrecognized blends at the same
redshift, the situation is only simple if the two objects have the same spectral energy distribution
(SED).  If they do not, then the composite object will correspond to some possibly strange SED,
which may not give a correct photo-$z$.  If the objects
are at different redshifts, it is unclear how the shear estimate should be interpreted (though the
case of large flux ratios or small redshift differences is simpler than the completely general
case).  The photometric redshift estimation is also complexified by the superposition of
SEDs at different redshifts, even for reasonably large flux ratios.  Unfortunately, the majority of
unrecognized blends will be at different redshifts (Kirkby et al.\ {\em in prep.}), except perhaps in the centers of
galaxy clusters.

The weak lensing community has recently come to confront the issue of blends more directly; this area requires more work, both
on the deblending algorithms and the post-deblender systematics quantification and
mitigation.  In the past, it was common practice to eliminate galaxies recognized as having nearby neighbors;
e.g., this was done in CFHTLenS\footnote{More specifically, galaxies with overlapping isophotes were
  rejected, while those that did not overlap very strongly and for which one galaxy could be masked
  without too much influence on the fit results for the other galaxy were retained.}
\citep{2013MNRAS.429.2858M}.  That approach does not help with unrecognized blends, and
can give a few-percent scale-dependent bias in shear-shear correlations due to the fact that close
pairs are more prevalent in high-density regions \citep{2011A&A...528A..51H} unless a weighting
scheme is used to mitigate that effect.  The early Dark Energy Survey Science Verification results
imposed cuts on recognized blends \citep{2016MNRAS.460.2245J} and ignored the issue of unrecognized
blends (while leaving a 5\% uncertainty on the multiplicative bias to cover this or other
uncorrected issues).  Their Year 1 (Y1) results included a more
sophisticated multi-object fitting strategy, with a careful study of the impact of blending using
simulations \citep{2017arXiv170801534S}.  In the HSC survey, which is deeper,
\citet{PipelinePaper:inprep} notes that 58\% of the objects 
detected in the HSC Wide survey are recognized blends.  As a result, rejecting blended objects is not a viable strategy, and estimation and
removal of blend-related systematics is necessary from the outset\footnote{A cut on the
  ``blendedness'' parameter in \citet{CatalogPaper:inprep} removes a very small
  fraction of objects, of order 1\%, that were dominated by spurious detections near
  very bright objects.  It does not affect the much larger fraction of genuinely
  blended real galaxy detections.}.  This was done using simulations that included a realistic level
of nearby structure around galaxies \citep{2017arXiv171000885M}.

For future surveys such as LSST, removing blended objects will not be a viable strategy due
to both their prevalence and the fact that a truly non-negligible fraction of them are unrecognized.
\citet{2016ApJ...816...11D} quantified the unrecognized blend population as exceeding $10$\% for a
survey like LSST, and investigated differences in its intrinsic ellipticity distribution, which is
relevant for weak lensing cosmology.   \citet{2013MNRAS.434.2121C} included blending when quantifying
the expected galaxy source number density and redshift distribution from LSST, and estimated the
impact of rejecting those blend systems recognized as seriously blended.  They note that the need to
reject these objects depends on our ability to quantify
and remove blending systematics (and that the threshold
for ``seriously'' blended depends on the deblending and measurement algorithms).

The combination of space-based imaging from Euclid and/or WFIRST with
LSST ground-based images has the potential to benefit LSST on this issue 
(e.g., \citealt{2015arXiv150107897J} and Rhodes et~al.\ {\em in prep.}).  One could imagine a separate round of joint pixel-level
analysis resulting in forced deblending for LSST based on higher-resolution space-based
imaging.  In the area of overlap between those surveys, the space-based data can substantially aid
in deblending mildly blended cases and in detecting a larger number of otherwise unrecognized
blends. It can also be used to learn about the impact of blending for 
LSST alone, and develop systematic error budgets for the entire LSST survey region.

More work is clearly needed to develop a framework for quantifying systematics in shear inference
and photometric redshifts due to both recognized and unrecognized blends for both ground- and
space-based surveys.  The impact of blending on photometric redshift estimation and inference of the correct redshift distribution for photometric
redshift-selected samples is particularly tricky, since the spectroscopic redshift selection and
failure rate for samples used for training and calibrating photometric redshifts may result in them
having different unrecognized blend rates than the general galaxy population.  Both improved deblenders
and systematics mitigation schemes would be beneficial.  For all surveys, algorithms that include color and other information in new ways will
likely be explored in the coming years \citep[e.g.,][]{2016A&A...589A...2J}; another example is the
secondary deblender used in crowded cluster regions for DES
\citep{2015PASP..127.1183Z}.  Unfortunately, well-established algorithms for crowded stellar fields
\citep[e.g.,][]{1987PASP...99..191S} are not relevant in this situation, and more algorithm
development is needed for the complex multi-galaxy and star-galaxy blend systems that
will dominate surveys like LSST.




\subsection{Image combination}\label{subsec:imagecomb}


Nearly all surveys used for weak lensing science have multiple
images at each point that must be combined to reach the full survey depth.  Taking
multiple exposures can be helpful to prevent excessive build-up of artifacts like cosmic rays or
saturated stars on any one image, to fill in the gaps between CCD boundaries or artifacts, and 
to build a Nyquist-sampled image out of multiple undersampled images. The
manner in which image combination is done is important for weak lensing science.  The primary
consideration that determines the available algorithms is whether the imaging is Nyquist
sampled or not, which is primarily a difference for ground vs.\ space-based imaging.  For
space-based imaging, the primary concern during the image combination stage is how to properly
reconstruct a Nyquist-sampled image, while for ground-based imaging, the individual images are
Nyquist sampled, and the main question is how to optimally combine information across exposures.

Most space-based instruments are (by design) not Nyquist sampled, but a wise choice of
dithering strategies enables reconstruction of a Nyquist-sampled image.  The need to reconstruct
a Nyquist-sampled image from multiple undersampled images should factor into survey strategy for
space-based weak lensing surveys; e.g., the expected rate of cosmic rays 
should be factored into the calculation of how many exposures are needed at each
point (to allow for periodic losses and still obtain a Nyquist-sampled image).  The first principled method
for combination of space-based telescope images was presented in \citet{1999PASP..111..227L}.  This
method used linear algebra to solve out the aliased Fourier modes given some sub-pixel
dither pattern, and reconstruct a Nyquist-sampled image.  
\citet{2011ApJ...741...46R} generalized this approach to address several
challenges: the fact that when dither patterns are comparable to the side of individual chips, the
different exposures at each point will experience different field distortions; the ability to handle
holes due to, e.g., cosmic rays; and the fact that the input PSFs for the different exposures may
differ.  Their method,
\texttt{IMCOM}\footnote{\texttt{https://github.com/barnabytprowe/imcom}}, has not yet been used for
survey data, but it may be the best approach to use for
surveys like WFIRST.  The commonly used MultiDrizzle method \citep{2002PASP..114..144F} carries out
interpolation on the individual non-Nyquist-sampled exposures.  This problematic method necessarily causes
stochastic aliasing of the PSF \citep{2007ApJS..172..203R}, which means that aliased modes are not fully removed, and the PSF of the
resulting coadded image may not even be constant across a galaxy image.  Later updates to the
MultiDrizzle method were designed to eliminate these high-frequency artifacts and convolution with
an interpolant kernel \citep{2011PASP..123..497F}.  Finally, multi-epoch model fitting is a valid
approach to image combination for undersampled data, and in some sense may be the most optimal
approach, provided that a good per-image PSF model is known.  Further investigation on this point is
needed to weigh the tradeoffs between these options.

For ground-based imaging, there are multiple image combination options.  The first is to use a coadded image for all
science, including PSF estimation.  Several challenges make
this approach sub-optimal.  These include the fact that depending on how coaddition is carried out,
the coadd may not even have a well-defined PSF at each point (for example, inverse variance
weighting that depends on the total flux, or use of a median for the
coadd).  The answer to this challenge is to generate the coadd in a principled way that results in
a single well-defined PSF at each point \citep[e.g.,][]{PipelinePaper:inprep}.  The primary
remaining systematic challenge for this method is that the PSF changes discontinuously wherever there
is a chip edge in any exposure contributing to the coadd at a given point.  Given the
typical stellar density in images used for weak lensing, it is very difficult to model
these small-scale changes in PSF.  There is an additional concern with
respect to statistical errors, since coaddition may effectively discard information
that is present in the best-seeing images.

The second option is to use a coadded image for measurements of object properties, but produce the
coadded PSF based on the appropriate weighted combination of the single-exposure PSF models.  This
``stack-fit'' approach was used for the DLS and HSC surveys
\citep{2013ApJ...765...74J,PipelinePaper:inprep}.  It enables the elimination of the primary
systematics concern
with the first approach mentioned above (estimation of the coadd PSF), while retaining concerns such as information loss from the best-seeing exposures.  There are also lower level concerns
about relative astrometry, which can behave like a blurring kernel in the coadded image.  If the
relative astrometric errors are well-characterized, they can be accounted for by including this
blurring in the coadded PSF, though in the HSC survey this was not necessary because the
astrometric errors were sufficiently small as to have a negligible effect on weak lensing
\citep{CatalogPaper:inprep}.  While modeling the PSF discontinuities is not a problem with this
approach, it is still the case that some fraction of objects will be lost due to their being so
close to the edge of a sensor in some exposure that the PSF cannot be modeled as effectively
constant across them.  This issue is more important as the number of contributing exposures in the
coadd increases.

Finally, it is possible to use a coadd for object detection and deblending, while measuring 
 shear and (optionally) galaxy photometry through simultaneous fitting to the
individual exposures.  This approach was proposed for LSST \citep{2008ASPC..394..107T}, and used for
DES Y1 \citep{2017arXiv170801533Z}, CFHTLenS \citep{2013MNRAS.429.2858M}, and KiDS.  In principle,
this allows for marginalization over the centroid positions in the individual exposures, which was used
in CFHTLenS to marginalize over relative astrometric errors.  Additional benefits include the fact
that the information in the best-seeing images can be preserved.  There are several limitations to
this approach.  First, some measurement methods (such as those based on measurement of moments) do not map to a combined
likelihood framework and therefore cannot
be used in a simultaneous fitting approach.  Second, this approach is
computationally intensive. When producing a best-fitting model, with $M$ iterations, fitting to a
coadd requires $M$ model convolutions with the PSF, while fitting to $N$ individual exposures with
different PSFs requires $N\times M$ model convolutions. For LSST, which will have hundreds
of exposures per object in the final dataset, this will require clever optimization to reduce
computational expense.  Alternatively, use of a fully Bayesian shear estimation method
\citep[e.g.,][]{2016MNRAS.459.4467B} that includes computation only of moments on each exposure would be a less
expensive way to combine information from all exposures; in this case, the complexity is moved into
the later shear inference step, and will not scale linearly with $N$.

Several ideas have been proposed to optimize the simultaneous fitting approach so that it is more
feasible for future surveys.  First, as in one of the
methods used in DES \citep{2015ascl.soft08008S}, representing the galaxy and PSF models as sums of Gaussians with different
scale radii can drastically speed up the calculations.  This could be done to produce an
initial guess at model parameters before switching to a model with full complexity.  Or it could
be done in combination with a technique like metacalibration, which will soak up the error induced
by model simplification in its estimate of the shear response (see Section~\ref{subsec:shearest} for
a discussion of metacalibration).  Second, it would be possible to fit for object properties using a
coadd to get an initial guess, and then tweak that model using fits to individual exposures.  Third, it may be
possible to use some hybrid of the simultaneous fitting and coadd approaches.  For example, if the
LSST exposures were split into 10 sets based on percentiles in the PSF size, and each of those sets
were coadded, then they could be used for simultaneous fitting with very little loss of information
in the best-seeing exposures. This would also alleviate the concern raised above with loss of
objects falling on PSF discontinuities, since fewer exposures would contribute to each coadd.
 Further investigation into the interplay between
information gain/loss and systematics for these approaches is needed in order to define a path
forward for the field (Sheldon et~al.\ {\em in prep.}).  

Finally, as suggested
above, there is some connection between methods used for image combination, deblending, photometry,
PSF and shear estimation; the best solution for image combination may depend on what is being done
for the other steps of the analysis process.

\subsection{Selection bias}\label{subsec:selbias}


Selection bias arises when quantities used to select or weight the galaxies entering the
lensing analysis depend on the galaxy shape.  Usually this dependence is implicit rather than
explicit, due to the details of image analysis algorithms or lensing magnification (which modifies
sizes and brightnesses in a way that correlates with the shear).  As a result, the probability of a galaxy entering the
sample (or its assigned weight) depends on its alignment with respect to
the shear or PSF anisotropy direction.  This violates the assumption that galaxy intrinsic
shapes are randomly oriented.  If the selection probability depends on the shear, there
will be a multiplicative bias, whereas if it depends on the PSF shape, there will be an additive
bias.

In the case of continuous quantities used for selection (such as galaxy size) causing
the bias, the bias is present only for galaxies 
near the boundary of the sample in the quantity used for selection.   In contrast, while weights
used to construct weighted averages are also continuous variables, biases
related to the weighting scheme used to take ensemble averages \citep[e.g.,][]{2017MNRAS.467.1627F}
may be present throughout the sample.  Finally, there are selection biases such as  avoiding
elongated structures (bad CCD columns) that can lead to
selection bias depending on how they are imposed \citep{2014MNRAS.440.1296H}.  Cuts on continuous
quantities and on regions like bad columns are appropriate to avoid certain systematic errors, but
if not applied with care, they can cause a selection bias.

There are several approaches to selection bias.  One is to simply avoid it: define detection
significances and apparent sizes compared to the PSF using round kernels, so
that the results are insensitive to galaxy shapes \citep{2016MNRAS.460.2245J}.  Another
way is to estimate its magnitude through an analytic formalism based on moments
\citep{2004MNRAS.353..529H,2005MNRAS.361.1287M}, and remove it. Its magnitude can be estimated using
simulations and then removed, if the simulations 
include all physical effects that induce selection biases.  For example, if the photometric
redshifts are coupled to the galaxy shape then there could be a selection bias that can only be
estimated using realistic multi-band simulations.   Moreover, these estimation methods must take
into account how selection bias varies based on the full range of observing conditions in the
survey.  Finally, self-calibration approaches to
shear estimation such as metacalibration (discussed in Sec.~\ref{subsec:shearest}) offer the
opportunity to directly estimate using only the real data, and remove selection bias. 

Selection bias plays an important role in defining useful null tests
\citep[e.g.,][]{2005MNRAS.361.1287M,2016MNRAS.460.2245J}.  For example, it is commonly suggested that
subsets of the galaxy sample should be used to carry out the lensing measurement, with the 
subsets giving consistent cosmology results in the absence of a systematic that depends on the
quantity being used to divide the sample.  However, the process of dividing up the sample can induce
selection biases that are of order 5-10\% depending on how this is done.

Until recently, selection biases have not attracted nearly the attention given to other shear biases
(Sec.~\ref{subsec:shearest}).  In future, more work is
needed to avoid selection biases from complex cuts related to galaxy blends, bad pixels, and other
selection criteria that cannot be mitigated as straightforwardly as S/N or size.


\subsection{Other aspects of the image processing}


Several other steps in the image processing can affect weak lensing 
besides those explicitly called out in previous subsections.  
  First, in order to detect and measure galaxy properties, the sky level must
be estimated and subtracted.  Errors in sky subtraction can cause coherent problems with 
object detection, photometry, and shear estimation near very bright objects -- bright stars or
collections of bright galaxies (e.g., in galaxy clusters).  A spurious sky gradient can induce a
spurious 
shear with respect to the location of the bright object causing the sky misestimation.  This
effect was identified and its impact on object detection, photometry, and shapes was
quantified in the SDSS \citep{2006ApJS..162...38A,2011ApJS..193...29A}.

Another relevant issue is star-galaxy separation.  There are two issues: the bright star
sample used to estimate the PSF can be contaminated by galaxies; and the faint galaxy sample
used to estimate lensing shear can be contaminated by (unsheared) stars.  For current surveys, we
have no evidence that star-galaxy separation algorithms are failing at problematic levels \citep[e.g., in
HSC,][]{PipelinePaper:inprep,CatalogPaper:inprep}. Given that more sophisticated methods for
star/galaxy classification have been proposed, for example using machine learning, there is clearly
room to improve to the level needed for upcoming surveys. 
Slightly more interesting issues with star-galaxy separation include binary stars contaminating the
galaxy sample \citep{2017MNRAS.465.1454H}.  In principle, these can be identified by looking for centroid offsets between different filters for highly
elongated objects, for those binaries in which the stars have 
different SEDs.

The primary astrometric concern for weak lensing is the accuracy of the relative astrometry between
different 
exposures for individual objects.  The relative astrometry must be well-understood in order to fully
understand the object measurements from simultaneous fitting and/or coaddition.  Systematics due to
errors in relative astrometry depend on exactly how the image combination is carried out; see
Section~\ref{subsec:imagecomb}.  For an example of how astrometric calibration was carried out for
the Dark Energy Survey, including correction for certain detector non-idealities (see
Section~\ref{subsec:detector}), see \citet{2017arXiv170609928B} and \citet{2017PASP..129g4503B}.
The astrometric calibration must include color terms to account for centroid shifts
from differential chromatic refraction (DCR) and other low-level effects. 

Finally, modeling of the noise in images is relevant to weak lensing measurements.  
Correlated pixel noise can arise due to 
low-level unresolved galaxies just below the detection threshold, methods used to combine
multiple exposures into a coadd (Section~\ref{subsec:imagecomb}), and pixel-level correction for
effects such as CTI (Section~\ref{subsec:detector}).  Since correlated noise means that detection
significances differ from the values one would naively assume given uncorrelated noise with the same
variance, and shear biases depend on the detection significance
\citep{2016MNRAS.457.3522G}, it is important to understand the noise correlations.

\subsection{Shear estimation}\label{subsec:shearest}


Since the initial detections of weak lensing shears around galaxy clusters in the 1990s, a large fraction of the weak lensing community's technical concern effort has
focused on the challenge of correcting galaxy shapes for the impact of the PSF so they can be
averaged to infer the lensing shear. In the past two decades, the field has moved from simple
methods based on correcting second moments of galaxy images for the moments of the PSF
\citep[e.g.,][]{1995ApJ...449..460K}, to a broader set of methods that include 
fitting parametric models \citep[see methods described in][]{2007MNRAS.376...13M},
to greater conceptual sophistication in how shear should be inferred \citep[see methods
described in][]{2015MNRAS.450.2963M}.  The community has set itself a series of blind
challenges
\citep{2006MNRAS.368.1323H,2007MNRAS.376...13M,2009AnApS...3....6B,2010MNRAS.405.2044B,2010arXiv1009.0779K,2012MNRAS.423.3163K,2014ApJS..212....5M,2015MNRAS.450.2963M}
aimed at benchmarking the performance of shear estimation methods in a common setting,
understanding the main challenges, and in the process developed an open-source, well-validated
 image simulation software package \citep[\texttt{GalSim}\footnote{\texttt{https://github.com/GalSim-developers/GalSim}};][]{2015A&C....10..121R}.

It is important to note that we do not care about galaxy shapes.  Indeed,
the concept of a single number characterizing the galaxy shape is not well-defined 
in the presence of ellipticity gradients
and irregular galaxy morphology.  Given those real physical effects, the measured shape
will depend on the radial weight function.  Moreover,
even for a galaxy with elliptical isophotes, shapes must be measured with weighted moments to reduce
noise, and the measured shape will depend on the shape of the weight function.   The assignment of ``shapes'' to
individual galaxies is effectively the assignment of a single estimate of the local shear from each
galaxy image, along with the assumption that the best ensemble estimator is the weighted mean of
the point shear estimators.  For this reason, comparison of
galaxy shapes measured with different algorithms or in different surveys is rarely useful, and 
ensemble shear statistics provide the only meaningful comparison.

Shear systematics are
often categorized into ``multiplicative'' or ``additive'' (in terms of what they do to ensemble shear
statistics). Additive systematics can have quite different scale dependence from
lensing shear correlations, depending on their physical origin.  In
Section~\ref{sec:obssys} I discuss methods for empirically identifying additive bias.  In contrast,
multiplicative bias cannot be easily identified through null 
tests; typically simulations are required.  While exact requirements vary depending on the details
of the survey and the assumptions made about the weak lensing analysis, typically the upcoming Stage
IV surveys require understanding of the shear calibration at the level of $\sim 2\times 10^{-3}$ in
order to avoid this systematic uncertainty dominating over the statistical uncertainties in the
measurement.  This is a factor of several smaller than the requirements for the measurements with
the full areas of ongoing surveys, and includes all sources of multiplicative calibration
uncertainties (e.g., PSF modeling errors), not just those due to the insufficiency of the PSF
correction method.

The past five years have seen a shift in how the field approaches shear estimation.
From the mid-1990s until $\sim$2011, the primary goal of weak lensers was
to find ways to estimate per-object shapes that, when averaged together, provided an unbiased
estimator for the ensemble shear.  During that time period, the typical
magnitude of shear calibration biases decreased by a factor of a few.   However, by that point it
was becoming increasingly obvious that the ``measure galaxy shapes and average 
them to get the shear'' approach has fundamental flaws from a mathematical
perspective.  An example is noise bias, wherein the maximum-likelihood estimate of per-galaxy shapes
at finite signal-to-noise is biased because noise changes the shape of the likelihood surface
\citep{2002AJ....123..583B,2004MNRAS.353..529H,2012MNRAS.427.2711K,2012MNRAS.424.2757M,2012MNRAS.425.1951R}.
Another example is model bias, which arises from the failure
of model assumptions to describe real galaxy light profiles \citep[e.g.,][]{2010MNRAS.404..458V,
  2010A&A...510A..75M}.  Selection bias (Sec.~\ref{subsec:selbias}) is another limitation of this
approach.

Some proposed solutions for model and
noise bias compete with each other:
increasing model complexity may decrease model bias, while increasing noise
bias due to the need to constrain additional degrees of freedom.  Shear estimation methods
based on measurements of per-galaxy shapes must balance 
these two considerations, with a finite amount of both biases in the ensemble shear
estimates.  Any method based
on the use of second moments to estimate shears cannot be completely independent of the details of
the galaxy light profiles, such as the overall galaxy morphology and presence of detailed
substructure \citep{2007MNRAS.380..229M,2010MNRAS.406.2793B,2011MNRAS.414.1047Z}.  Nor is noise bias
avoidable: given the large intrinsic galaxy shape dispersion, lensing measurements must include
galaxies down to relatively low signal-to-noise detections to achieve a reasonable overall
signal-to-noise in the ensemble shear statistics.

Given the recent understanding of this situation, the community has sought other approaches to reliable
ensemble shear estimation.  There are four general classes of approach, some of which are compatible
with each other.

 \textbf{Image simulations to estimate and remove calibration biases:} Image
  simulations enable the estimation of biases in the shear signal due to the
  intrinsic limitations of the adopted shear estimation method.  Given that shear biases depend on
  detailed galaxy morphologies (beyond second moments) and on the PSFs, there has been a move
  towards ever greater realism in the image simulations used by ongoing surveys
  \citep[e.g.,][]{2017arXiv170801533Z}.  For example, several works have argued that one must
  include nearby structure around the galaxies in order to accurately
  predict shear biases due to nearby objects and unrecognized blends
  \citep{2015MNRAS.449..685H,2017MNRAS.468.3295H,2017arXiv171000885M}, must account for variation of these biases with
  observing conditions across the survey, and have identified other key factors in image
  simulations for shear calibration.  This approach will be challenging to take to the
  limit of future surveys, given our limited knowledge of galaxies, although the survey data itself
  provides a form of sanity check on the accuracy of the simulations and perhaps could enable an
  interative process to improve the simulations \citep[see, e.g., the \texttt{sfit}
  method:][]{2015MNRAS.450.2963M}. To ensure that the statistical error on the derived bias corrections is a
  subdominant part of the overall error budget, it is necessary to simulate many more galaxies than
  exist in the survey itself.  Moreover, use of calibrations as the sole way of estimating and
  removing shear biases does not provide an independent cross-check on the results (unlike, e.g.,
  using one of the methods of calibrating the shear below, and then using simulations to validate
  that method as a cross-check).  

\textbf{Self-calibration:} Recently, methods have been devised to calibrate ensemble shear
  statistics based on manipulations of
  the real images 
  \citep[``metacalibration'':][]{2017arXiv170202600H,2017ApJ...841...24S}.  Metacalibration provides
  a way to determine the response of an ensemble shear estimator for the real galaxy population
  in the data.  This potentially enables direct removal of selection biases, depending
  on what stage of the image processing metacalibration is inserted into.  The fact that it does not
  require assumptions about galaxy morphology is a clear virtue of this approach over image
  simulations.  

\textbf{CMB lensing:} Since CMB lensing has very different observational systematics
  and a perfectly known source redshift, it is an attractive method for testing galaxy lensing
  \citep{2012ApJ...759...32V,2013arXiv1311.2338D}.  Comparison of galaxy lensing with CMB lensing
  should not be thought of as a test of galaxy shear estimation, but rather of a
  combination of shear estimation and photometric redshift biases that both modify the
  lensing signal amplitude.  When measuring cross-correlations between CMB and galaxy lensing, intrinsic
  alignments (Section~\ref{subsec:ia}) are a contaminant that must be modeled
  \citep{2014MNRAS.443L.119H,2014PhRvD..89f3528T,2015MNRAS.453..682C}, unlike for correlations
  between the galaxy
  positions and CMB lensing.  While current imaging and CMB surveys can only provide a $\sim 10$\%-level calibration \citep[e.g.,][]{2016MNRAS.461.4099B,2016MNRAS.460..434H,2017MNRAS.464.2120S},
  the situation will improve with future galaxy and CMB surveys \citep{2016arXiv161002743A}. As
  demonstrated by \citet{2017PhRvD..95l3512S} 
  with forecasts that include systematic uncertainties, this method is unlikely to
  constrain the lensing signal amplitude at the level of precision needed to avoid a survey
  like LSST being systematics-dominated.  However, it provides a valuable independent cross-check on
  the other methods of shear calibration in this list. 
 Moreover, it may allow the shear calibration at high redshift to be
  constrained at the level needed for LSST, which is helpful because that is likely the regime
  with the most uncertainty on how to simulate the galaxy population.

\textbf{Paradigm shift:} The final approach described here, and perhaps the most principled
  one, is that since the meaning of per-galaxy shapes is questionable (given ellipticity gradients
  and other effects) and the approach of averaging them is fundamentally mathematically flawed, 
  we should stop doing this.  Instead, we should directly infer ensemble shear statistics in a
  way that avoids these assumptions, using the actual posterior shear estimate from each galaxy
  without assuming that an ellipticity is an unbiased proxy for it. 
 \citet{2015ApJ...807...87S} explored hierarchical
  inference of the shear, which involves parametric model fits that are then used to infer ensemble
  shear given a prior; this appeared promising, but requires further development due to the computational expense of the approach.  \citet{2014MNRAS.438.1880B} presented a Fourier-space Bayesian shear
  estimation method that does not involve averaging galaxy shapes, and should not be susceptible to
  either model or noise bias.  This method involves measurements of moments in Fourier space for
  the galaxy sample to be used, and the construction of a prior for what the unlensed distribution of
  moments looks like using a deep subset of the same survey. Together, these can be used to infer the
  ensemble shears. The method was 
  developed in subsequent work \citep{2016MNRAS.459.4467B} to bring it closer to a practical shear
  estimator for use in real data, and the self-consistent modeling of photometric redshifts,
  selection biases, and measurements in multi-epoch data seems possible in principle.
  While work is needed to fully demonstrate the utility of these methods that overthrow the
  traditional paradigm in real data, particularly the extension to unrecognized blends that are not
  at the same redshift, 
  their mathematical justification is unquestionable.  A first application of the method from \citet{2016MNRAS.459.4467B} to the HSC survey will be
  presented in Armstrong et~al.\ {\em in prep.}

It is becoming increasingly common for weak lensing surveys to use two shear estimation
methods with different assumptions \citep[e.g., DES Y1 results in][]{2017arXiv170801533Z}, relying on the comparison to
provide some support for the reliability of survey results.  A combination of a ``traditional
method'' calibrated using method (1) or (2), compared with a method in class (4), and an
external calibrator like CMB lensing (3), may be necessary to fully justify a belief in the
results of Stage-IV lensing surveys at the level of their statistical errors (i.e., without addition
of a substantial systematic error budget).


\subsection{Photometric redshifts}\label{subsec:pz}


In this section, I discuss the calculation of photometric redshifts\footnote{Many methods produce a
photometric redshift posterior probability $p(z)$ rather than a single point estimate.  I will nonetheless refer to these indiscriminately as photometric redshifts or
photo-$z$'s.}, or photo-$z$'s.  For weak
lensing, what primarily matters is the ability to infer the true redshift distribution for
a photo-$z$-selected sample of galaxies.  In other words, there are strong requirements on our 
  knowledge of the photometric redshift errors.  I will discuss methods to accurately calibrate the redshift
distributions in Section~\ref{subsec:nz}, while in this section I focus on 
the photometric redshift estimation itself.

The first step in calculation of photometric redshifts is to measure the input data, which
most commonly consists of flux measurements in several photometric bands\footnote{Several
  papers have suggested using additional morphological information, such as sizes and shapes
  \citep[e.g., most recently,][]{2017arXiv170703169S}.  Given that these correlate with lensing shear, it is unclear what the impact of this
  would be for cosmology analyses.  For example, if the photo-$z$ errors become systematically
  correlated with the lensing shear, this could be problematic to correct.}. For example, consider a galaxy without any color gradients.  If
the PSF is the same in all bands, aperture photometry might be a perfectly reasonable way to get
stable color estimates.  Given that the PSF typically differs between the bands, aperture photometry
will not give stable color estimates unless the aperture size is large compared to the
PSF in the band with the worst seeing, which would result in quite low S/N.  As a result, typically some form of PSF-matched aperture photometry
\citep[e.g.,][]{2012MNRAS.421.2355H} or
forced model photometry (with the same model used in each band; e.g., \citealt{2017arXiv170405988T})
gives better results.  More generally, the multi-band photometry must measure light from the same
physical area of a galaxy to properly estimate the SED, even if that light comes from a subset of
the galaxy (chosen consistently across the bands).

Ideally, these
measurements should be made in a way that reduces sensitivity to systematics such as Galactic
extinction, seeing, and other observational or astrophysical effects with coherent
patterns on the sky. 
There are some low-level systematics to consider in calculation of the
photometry, e.g., variation of the bandpasses across the field-of-view and photometric calibration
across the survey including color effects \citep{2016AJ....151..157L,2017arXiv170601542B}.
For these and other spatially-varying issues such as Galactic extinction, it is not generally the RMS
error that is relevant, but rather the spatial correlation function of the errors, which will
determine the scales on which the measured two-point correlations will show signatures of these
systematics.  See \citet{2009ApJ...690.1236I} for a discussion of technical 
considerations such as uncertainty in photometry/filter curves.

There are several classes of photometric redshift methods; see \citet{2010A&A...523A..31H} for a
summary of many methods, and \citet{2017arXiv170405988T} and
\citet{2014MNRAS.445.1482S} for the methods used for HSC and DES Science Verification, respectively.  The two main classes of methods are (1) template-fitting methods, which rely on a set of
templates for galaxy SEDs that are used to predict the galaxy photometry as a function of redshift,
and can be compared with the observed photometry \citep[for a summary, see][]{2006A&A...457..841I}; and (2) machine learning methods, which empirically
learn the relationship between photometry and redshift based on a training sample.  The key issues
for template-fitting methods are insufficiency of the templates to accurately describe the full span of the real data, while the
key issues for machine learning methods are the difficulty in generalizing to samples that do not
look like the training data.  Both of these limitations would be 
eliminated if we had a very large, perfectly representative spectroscopic training
sample -- which highlights the fact that the primary limitation for modern photometric
redshift methods is the insufficiency and/or non-representativeness of spectroscopic redshift
samples to the depth of the lensing surveys \citep{2015APh....63...81N}.  

There are multiple problems with existing spectroscopic samples.  Some regions of color and magnitude space 
are not well-covered by spectroscopic redshift samples, particularly at the faint end.  In
principle, reweighting schemes \citep{2008MNRAS.390..118L,2009MNRAS.396.2379C} could mitigate
this limitation when training and/or calibrating photometric redshifts, as long as all regions of
color and magnitude space have some objects.  Unfortunately, this solution may not work because it is not obvious that
spectroscopic redshift successes and failures at fixed color and magnitude have the same redshift
distribution.  This problem is much harder to detect without e.g.\ obtaining spectra
from a different spectrograph that has a different range of wavelengths and sensitivity.  
Also, since the galaxy samples used for weak lensing have additional selection criteria imposed besides
cuts on color and magnitude, 
it may be necessary to consider this higher dimensional space when training and calibrating
photometric redshifts for lensing \citep[e.g.,][]{2017arXiv170801532H,2017arXiv170600427M}.  Trying to match
this higher dimensional space is challenging given the limited size of current
deep spectroscopic samples.  Finally, it is possible that some selection criteria used for targeting galaxies
for spectra can induce additional non-negligible biases in the redshift distribution, which is
problematic when using those redshifts for spectroscopic training and calibration
\citep[e.g.,][]{2017MNRAS.468..769G}.

One outstanding problem in the field is photometric redshift training in the presence of unrecognized blends (see Section~\ref{subsec:deblending}).  This is a
non-trivial problem that requires additional attention from the field as we move towards
deeper surveys.  One approach may be to ignore this issue in training, and fold it into the
catastrophic failure rate when calibrating the photo-$z$; this places greater demands on the
calibration strategy.  In addition, the existence of shear selection biases induced by photo-$z$
complicates the analysis of tomographic shear correlations; see \citet{2017arXiv170801538T} for a recent example with
mitigation schemes.

\subsection{Masks and survey geometry}


Describing the survey coverage requires a way to describe the exact location
of its boundaries -- not just edges but also internal boundaries due to e.g.\ masking bright stars --
and the spatial dependence of quantities that determine systematic errors and/or galaxy number
densities, such as the depth, PSF size, etc.  Several software frameworks have been
developed to describe survey geometry, typically with some flexible hierarchical description of
geometry.  These include Healpix\footnote{\texttt{https://github.com/healpy/healpy}}
\citep{2002ASPC..281..107G}, Mangle \citep{2012ascl.soft02005S}, and STOMP
\citep{2007ASPC..382...85S}.

There are several places where these descriptions are needed.  First, maps of the spatial dependence
of systematics can be correlated against quantities of scientific interest (e.g., photo-$z$'s or shear
estimates) to identify which systematics are most relevant and need further improvement.
An example of map-level systematics investigation in the HSC survey was carried out by
\citet{Oguri:inprep}. Second, coverage maps can be useful to generate mock observations that have
the same coverage as the real data.  Since survey boundaries can lead to selection biases and to
leakage between E and B-mode power, mock catalogs with the same boundaries can be valuable for
systematics investigations. Finally, the optimal estimator for galaxy-galaxy correlations
\citep{1993ApJ...412...64L} and galaxy-shear correlations \citep{2017MNRAS.471.3827S} requires
random points with the same spatial coverage as the real galaxies (but with correlation
function equal to zero).  This need arises because the optimal estimator involves correlation of the
overdensity rather than the density itself, so in each case where the galaxy field is used, a random
field is needed also.  Moreover, for galaxy-shear correlations, the subtraction of shear around
random points not only produces a more optimal estimator, but is useful for subtraction of
systematics \citep{2005MNRAS.361.1287M} if the number density-dependence on
systematics-generating quantities is faithfully reproduced in the random sample
\citep{2013MNRAS.432.1544M}. \citet{2015MNRAS.454.3121M} have demonstrated the impact of systematic
variation of galaxy number densities with observational parameters such as depth, extinction, and so on, and
the need to model these dependencies beyond linear order to accurately estimate angular correlations
from large imaging surveys.  This is relevant both for galaxy-galaxy and galaxy-shear correlations
that go into a cosmological weak lensing analysis.

The above statements about the need to faithfully reproduce survey boundaries and the dependence of
galaxy density on observational conditions is related to arguments made in the literature about the
so-called ``boost factor'' that accounts for the contamination due to (unlensed) physically-associated galaxies
used as sources in galaxy-galaxy or cluster-galaxy lensing measurements.  This idea was introduced
by \citealt{2004AJ....127.2544S}.  The difficulty in using this formalism for small-area surveys
in practice was presented by \citet{2017arXiv170600427M} and \citet{2017MNRAS.469.4899M}, with an
alternative formalism involving explicit modeling of the smooth redshift distribution and the
contribution from physically-associated galaxies given by \citet{2014MNRAS.442.1507G}.  An
additional complication is the need to trace the possible difficulties detecting source galaxies in
high-density regions (e.g., in galaxy clusters, due to the obscuration of background galaxies
by foregrounds; \citealt{2015MNRAS.449.1259S}).

\section{FROM CATALOGS TO SCIENCE}\label{sec:cat2science}


This section covers the steps in a weak lensing analysis from catalogs to cosmological parameters.
It is in this phase of the analysis that we must include steps for mitigation of astrophysical
uncertainties and any residual observational systematics.

\subsection{Estimators}\label{subsec:estimators}


Here I assume the availability of a set of galaxy positions on the sky, per-object shear
estimates defined as in Sec.~\ref{sec:intro}, and photometric redshifts. The estimator for the
reduced shear will be denoted 
$\hat{g}$.  
Typically the coordinate system for $\hat{g}$ is defined such that positive $\hat{g}_1$
corresponds to an East-West or North-South elongation, while $\hat{g}_2$ is defined at $45^\circ$
with respect to that axis.  The focus of this section is how to combine these quantities
and measure statistics that are cleanly related to the matter distribution.

For shear-shear correlations, galaxies are divided into tomographic bins based on the photometric
redshifts.  Pairs of galaxies are identified, and their separation on the sky is calculated,
including the angle with respect to the sky coordinate axes: separation on the sky $|\theta|$ and
polar angle $\phi$.  For each pair, the relevant shear components are tangential $\hat{g}_+$ and
cross $\hat{g}_\times$, with the convention that tangential shear around overdensities results
in $\langle \hat{g}_+\rangle >0$ and radial shear around underdensities gives $\langle
\hat{g}_+\rangle <0$.  For one of the galaxies with shape $\hat{g}$, we obtain
\begin{align}
\hat{g}_+ &= -\text{Real}[\hat{g} \exp{(-2\mathrm{i}\phi)}]\\
\hat{g}_\times &= -\text{Imag}[\hat{g} \exp{(-2\mathrm{i}\phi)}].
\end{align}
The estimator for the shear correlation functions $\xi_\pm$ in that tomographic bin is \citep{2002A&A...396....1S}
\begin{equation}
\hat{\xi}_\pm(\theta) = \langle \hat{g}_+\hat{g}_+\rangle(\theta) \pm \langle \hat{g}_\times\hat{g}_\times\rangle(\theta)
\end{equation}
with $\langle \hat{g}_+\hat{g}_\times\rangle=0$ due to parity symmetry, and the averages being
weighted averages (typically inverse variance weighting, including the intrinsic shape
noise and measurement error).  This realistic estimator is insensitive to survey 
masks and boundaries.  The theoretical prediction for $\xi_\pm(\theta)$ can be
derived as Hankel transforms of the convergence power spectrum,
\beq
\xi_\pm(\theta) = \int \frac{\ell\,\mathrm{d}\ell}{2\pi} J_{0/4}(\ell\theta) [P^\text{(E)}_\kappa
(\ell) \pm P^{(\text{B})}_\kappa (\ell)].
\eeq
To lowest order, lensing produces only E-mode power (a pure gradient field), but there are low-level physical effects that
cause B modes (corresponding to a curl component; see Section~\ref{subsec:theory}).  Certain systematics can manifest as mixes of
E and B modes, and detection of B-mode power is one way to identify those systematics; however, not
all systematics produce B modes.

Since lensing produces primarily E-mode power, and the power estimated in each
$\ell$ bin should be roughly independent, there is interest in directly
estimating the power spectrum.  However, the most naive way of doing so involves measuring shear
correlations over all scales, and in practice, the lack of pairs on small scales and the finite
sizes of lensing surveys leads to a mixing of E and B modes
\citep{2006A&A...457...15K}. There are so-called pseudo-power spectrum estimators
\citep[e.g.,][]{2011MNRAS.412...65H} that aim to mitigate this effect and enable direct estimation
of the power spectrum.  There are additional configuration-space estimators, and estimators that
combine the estimated $\hat{\xi}_\pm(\theta)$ with various filters in ways that are meant to be more
optimal
\citep[e.g.,][]{2012A&A...542A.122A}.
No matter what estimator is used, models for systematic uncertainties must be
re-expressed in terms of those estimators in order to marginalize over and remove the
uncertainties.

The above discussion was focused on shear-shear correlation functions.  However, as mentioned
previously, the canonical weak lensing analysis for future surveys will include galaxy-shear and
galaxy-galaxy correlations, defined within tomographic bins in analogous ways.  When constructing
these estimators using the galaxy overdensity field, they have contributions
from both clustering and magnification. \citet{2009ApJ...695..652B} presents the relationship
between empirical estimators of these three two-point correlation functions and the
underlying theoretical quantities: lensing shear, magnification, and galaxy overdensity.

When choosing estimators for the galaxy-shear and galaxy-galaxy correlations, there are
different philosophical approaches.  On small scales, these correlations
depend on how galaxies populate dark matter halos.  One family of
estimators removes the small-scale information to avoid systematic uncertainty in cosmological constraints
due to astrophysical details \citep[e.g.,][]{2010PhRvD..81f3531B}.
Other approaches are to include the small scales, 
build models with nuisance astrophysical parameters, and
marginalize over them \citep[e.g.,][]{2006ApJ...652...26Y,2013MNRAS.430..725V}.  The choice of which
type of estimator to use depends on the users optimism in their ability to describe these
astrophysical uncertainties with sufficient realism to avoid substantial systematic errors while
using a simple model. 

Data compression from shear correlations or power spectra may be possible and even desirable.  The
number of data points in the estimator places serious demands on covariance matrix estimation (Section~\ref{subsec:covariances}) and the
cosmological parameter inference method (Section~\ref{subsec:inference}).  For that reason,
investigation of data compression methods such as those recently proposed for galaxy power spectra may be
beneficial \citep{2017arXiv170903600G}.

Finally, for the case of shear-shear correlations, a 3D lensing approach that avoids the need for
tomographic binning has been proposed and used in real data
\citep{2009MNRAS.399...48S,2011MNRAS.413.2923K,2014MNRAS.442.1326K}.  However, future work is needed
on how to use this in a joint analysis with galaxy-shear and galaxy-galaxy correlations, and
properly marginalize over systematics.

\subsection{Redshift distributions and bins}\label{subsec:nz}


Section~\ref{subsec:pz} described photometric redshifts, defined either as point
estimates or posterior probability distributions $p(z)$. This section will explain how they are used for science.  A variety of
schemes exist for dividing galaxies into tomographic bins, e.g., based on division of the sample
using the point photo-$z$ estimates.  Determination of the true ensemble redshift distribution\footnote{While it is tempting to
stack the per-object $p(z)$, which is a mathematically acceptable approach to using spectroscopic
redshifts, stacking per-object $p(z)$ violates the definitions of 
probability (Malz et~al.\ {\em in prep.}).  It is nonetheless often done.}, or $N(z)$, is critical
for cosmological analyses.  To
lowest order, weak lensing is primarily sensitive to the mean redshift and the width of the redshift
distribution of 
each tomographic bin \citep{2007MNRAS.381.1018A}; this fact is often used to motivate how
nuisance parameters for redshift uncertainty are included in the cosmological analysis
\citep[e.g.,][]{2017arXiv170801530D}.  The inclusion of catastrophic
photometric redshift errors complicates this issue \citep{2010ApJ...720.1351H}.

In general, spectroscopic redshifts are needed for photo-$z$ training (Sec.~\ref{subsec:pz}) and
calibration, where the type of redshift samples needed for these purposes differs
\citep{2015APh....63...81N}.  The typical required redshift sample size is of order
$10^5$ in order to reduce the systematic uncertainty on mean redshifts in tomographic bins to the
$\sim 10^{-3}$ level that is needed to avoid Stage IV surveys being systematically biased at a level
exceeding the statistical uncertainties.  There are
two methods for using spectroscopic redshifts to calibrate the $N(z)$ of photometric redshift
samples.  The first is to reweight the spectroscopic redshifts to match the observed properties of
the photometric sample  \citep{2008MNRAS.390..118L} and directly infer the $N(z)$, though there is some debate as to which sample
properties should be used for that reweighting \citep[see,
e.g.,][]{2017arXiv170600427M}.  Previous studies have explored the 
spectroscopic redshift sample size needed for direct calibration of $N(z)$, without
\citep{2008ApJ...682...39M} and with 
catastrophic errors \citep{2009ApJ...699..958S,2010MNRAS.401.1399B}.  Generically, this method requires a spectroscopic redshift sample
that covers all of the photometric color and magnitude space (not necessarily evenly, with reweighting
accounting for the non-representativeness of the spectroscopic redshift sample). 
An additional assumption is that the $N(z)$ 
at fixed color and magnitude is the same for spectroscopic successes and failures, which is likely
incorrect at some level.  The resulting systematic uncertainty is difficult to estimate and is
often ignored for current datasets.  \citet{2017arXiv170108748B} proposes a framework for exploring
this assumption for shallow surveys.  The extension of this test to deeper surveys (where
degeneracies between high and low redshift may be more important in determining spectroscopic
success) is of critical importance for future surveys that wish to rely on spectroscopic reweighting
to determine the $N(z)$ of photometric redshift samples.

The second method is to use the cross-correlation between the photometric redshift sample and some
non-representative spectroscopic redshift sample 
covering the full redshift range of the photometric sample with large enough area and sampling rate to allow the
clustering cross-correlation to be well-determined. 
Several variations on the cross-correlation or clustering redshift method have been proposed
\citep{2008ApJ...684...88N,2010MNRAS.408.1168B,2013MNRAS.433.2857M,2013arXiv1303.4722M,2013MNRAS.431.3307S}.
Differences between them include the choice of scales to use (purely linear bias scales, or small
scales as well); the method of modeling the galaxy bias for the photometric sample; and the
corrections for magnification bias, which induces nonzero correlations between galaxies in bins that
truly are
separated in redshift.  Recent results using this approach include
\citet{2016MNRAS.463.3737C,2017MNRAS.467.3576M,2017MNRAS.465.4118J}.

Because of the different assumptions behind these two methods, 
DES and KiDS used both to calibrate their $N(z)$
\citep{2017arXiv170801532H,2017MNRAS.465.1454H}, though for DES they used subsets of luminous red
galaxies with
high-quality photo-$z$'s rather than spectroscopic redshifts when carrying out the
cross-correlation analysis, and also used high-quality 30-band COSMOS photo-$z$'s for the direct
$N(z)$ calibration.  It seems likely that in
future, both methods will continue to be
used so as to have a cross-check on the resulting calibrated $N(z)$.

The needs for additional spectroscopic redshift samples for photo-$z$ training and calibration for
future surveys is summarized in  \citet{2015APh....63...81N}.   Techniques have been proposed for how
to identify the regions of color/magnitude space that should be targeted to fill in missing regions
of parameter space (including self-organizing maps, \citealt{2015ApJ...813...53M}, which were used for
targeting a new spectroscopic survey in \citealt{2017ApJ...841..111M}).

Finally, the difficulty in calibrating $N(z)$ is connected to the exact analysis
being done.  Use of galaxy-shear, galaxy-galaxy, and shear-shear correlations together may result in
less stringent needs for spectroscopic redshift calibration samples, while the need to jointly model
intrinsic alignments may result in more stringent requirements for how well we understand
photometric redshift uncertainties \citep{2009A&A...507..105J}.  An additional issue of relevance
especially for deep ground-based surveys is the role of blending systematics
(Section~\ref{subsec:deblending}), which have the potential to increase the catastrophic photometric
redshift error rate.  Since small-area spectroscopic redshift samples may have targeting criteria that
avoid obvious blends, the impact of blending on photometric redshift errors may
need to be assessed primarily through the cross-correlation method.

\subsection{Theoretical predictions}\label{subsec:theory}


To constrain cosmology with weak lensing measurements, theoretical predictions with an accuracy of
$\sim$1\% over a wide range of scales and cosmological parameters are needed
\citep[e.g.,][]{2005APh....23..369H}.  To interpret the shear-shear correlations alone, predictions
for the distribution of dark matter are needed, while joint interpretation with galaxy-shear
and galaxy-galaxy correlations requires a way of describing the distribution of galaxies.

Weak lensing measurements typically go quite far into the nonlinear regime, so an accurate
description of the nonlinear matter power spectrum is required.  This description can
come from large suites of $N$-body simulations with many values of cosmological parameters, and some
manner of interpolating between the values of parameters for which simulations were generated.  One option is to use simulations to calibrate a fitting
formula \citep[e.g., halofit:][]{2012ApJ...761..152T}.  Another approach is to use an emulator, such
as \citet{2014ApJ...780..111H}, which interpolates over cosmological parameter space using Gaussian processes.  While fitting formulae
and emulators have tremendous value in enabling fast, accurate calculations of the matter power
spectrum, using simulations directly can help (a) enable inclusion of
physical effects that might be difficult to incorporate through an analytic approach and which
depend on cosmology, such as density-dependent selection effects
\citep[e.g.,][]{2011A&A...528A..51H}; (b) allow for joint modeling with galaxy correlations; and (c)
include higher-order theoretical nuances.

When modeling the galaxy-shear and galaxy-galaxy correlations, the simplest assumption to make is
that the galaxy bias is linear (galaxy and matter overdensities are related as
$\delta_g = b \,\delta$) and that the galaxy and matter overdensities are perfectly correlated,
$r_\text{cc}= P_\text{gm}/\sqrt{P_\text{gg} P_\text{mm}}=1$.  These simple assumptions are valid at 
large separations, and fail for a variety of reasons on small scales. 
They were 
used in the joint analyses of the three galaxy and shear auto- and cross-correlations from DES
and KiDS \citep{2017arXiv170801530D,2017arXiv170605004V}.  In DES, to avoid sensitivity to
systematics from the linear bias assumption, the choice was made to limit the
analysis to relatively large scales, $>8$ and $>12h^{-1}$Mpc for galaxy-galaxy and galaxy-shear
correlations, respectively.  In both cases, tests were carried out to assess the sensitivity of the
results to this assumption.  

For future surveys, the measurements will have sufficient
signal-to-noise that it will be necessary to adopt more realistic models.  One option is a
perturbation theory-based model for $b(k)$ and $r_\text{cc}(k)$
\citep[e.g.,][]{2010PhRvD..81f3531B}, which has been used for a galaxy-shear and
galaxy-galaxy joint analysis in SDSS \citep{2013MNRAS.432.1544M}.  Another option is a halo model
approach, which provides a numerical description for the galaxy-matter and galaxy-galaxy
correlations based on how galaxies populate dark matter halos
\citep[e.g.,][]{2006ApJ...652...26Y,2013MNRAS.430..725V}, and which has been used in practice for
interpretation of BOSS galaxy lensing and clustering \citep{2015ApJ...806....2M}.  As mentioned in
Section~\ref{subsec:estimators}, the choice of the estimator to use for the measurement is related
to the question of how the modeling is to be done, because some model descriptions can go to smaller
scales than others.  In addition, it may be necessary to consider how higher-order complexities like
assembly bias (wherein the galaxy bias depends on more than just the mass) complicates the joint
analysis of galaxy and shear correlations, specifically assumptions about $r_\text{cc}$ and its
scale dependence.  Preliminary steps towards understanding this issue
have already been made \citep[e.g.,][]{2016arXiv160102693M}.  For future surveys, 
calibration of how $b(k)$ and $r_\text{cc}(k)$ are modeled against realistic mock galaxy
catalogs will be crucial for choosing what range of scales can be used and ensuring accurate
cosmological constraints.

For the prediction of projected lensing statistics, there are a number of low-level theoretical
issues that have not been relevant for past and ongoing lensing surveys, but which may require attention in upcoming surveys.  These include the distinction between shear and reduced shear, the
impact of several approximations (flat sky, Born, Limber, linearized gravity, and Hankel transform)
and higher order lensing terms, lens-lens coupling, and source clustering-induced B modes 
\citep{1998A&A...338..375B,2002A&A...389..729S,2006PhRvD..73b3009D,2009A&A...499...31H,2010PhRvD..81h3002B,2010A&A...523A..28K,2017arXiv170706640G,2017arXiv170205301K,2017MNRAS.469.2737K,2017JCAP...05..014L,2017PhRvD..95l3503P}.
Fast methods have been developed for ray-tracing through $N$-body simulations
\citep{2016JCAP...05..001B} to avoid some of these approximations, and to incorporate some of the
second-order effects \citep{2013MNRAS.435..115B}. These effects can
enter in different ways to the galaxy-shear correlations, e.g., because of lensing deflections
modifying observed positions.

\subsection{Intrinsic alignments}\label{subsec:ia}

\begin{figure}
\begin{center}
\includegraphics[width=\linewidth]{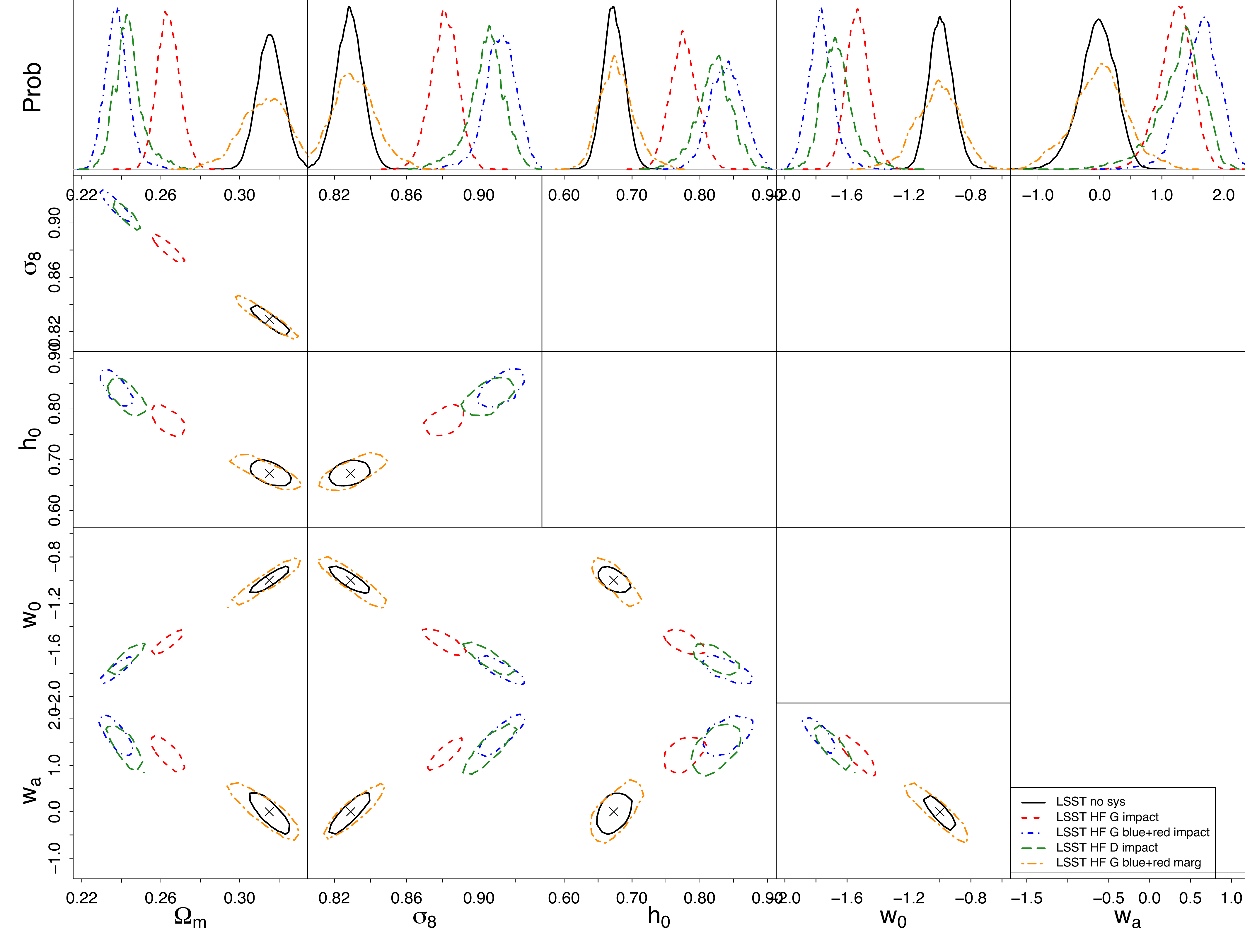}
\end{center}
\caption{An illustration of the impact of intrinsic alignments on cosmological parameter constraints
  with weak lensing in LSST, using shear-shear correlations only (from \citealt{2016MNRAS.456..207K}).
The bottom left triangle shows the 2D contours for cosmological parameters, where the black curve
shows the case of no intrinsic alignments; red, green, and blue curves show intrinsic alignments
predictions with different ways of modeling the alignments of blue and red galaxy populations and
their luminosity evolution; and the orange curve shows how the constraints become less tight when
marginalizing over the intrinsic alignments.  The top row shows the posterior probabilities for each
of the cosmological parameters.  Clearly the biases without marginalization are
unacceptably large.
}
\label{fig6}
\end{figure}
Since recent reviews have covered the physics of intrinsic alignments, their impact on
weak lensing cosmology, theoretical models, and observations 
\citep{2015SSRv..193....1J,2015SSRv..193..139K,2015SSRv..193...67K,2015PhR...558....1T}, this
section will
be brief.

Intrinsic alignments, the coherent alignment of galaxy shapes due to physical effects other than lensing, are
a major theoretical uncertainty for weak lensing \citep{2000ApJ...545..561C, 2000MNRAS.319..649H,
  2000ApJ...532L...5L,2001ApJ...555..106L, 2001MNRAS.320L...7C}.  Since lensing analyses must assume
that coherent shape alignments are due to lensing, intrinsic alignments can contaminate lensing
analyses. 
Theoretical prediction of intrinsic alignments is difficult because, while simulations and analytic models
show that dark matter halos have intrinsic alignments in the $\Lambda$CDM paradigm, the question of
whether observed galaxy shapes (the baryonic components) also show such alignments cannot be
answered with $N$-body simulations.  Depending on assumptions made about the galaxy population and
the alignments of its baryonic components with the underlying matter field, the predicted level of
alignments can vary by orders of magnitude \citep[e.g.,][]{2006MNRAS.371..750H}.  However,
observations have substantially narrowed this wide variation by placing constraints on the
large-scale alignment model for red galaxies, and (so far) null detections of large-scale shape
alignments for blue galaxies.  Since red galaxies exhibit alignments consistent with their
shapes being aligned with the shapes of the inner regions of their halos, high-resolution
$N$-body simulations may indeed be populated with red galaxies that have realistic alignments with
the underlying matter density field \citep[e.g.,][]{2012JCAP...05..030S}.  Recent work has also
included comparison of measured galaxy alignments with high-resolution, large-volume hydrodynamic
simulations that include the physics of galaxy formation
\citep{2015MNRAS.454.3328V,2016MNRAS.462.2668T,2017MNRAS.472.1163C,2017MNRAS.468..790H}; the
simulations broadly reproduce 
many of the alignment trends seen in real data, but not all of them.

Initial efforts to remove intrinsic alignments from lensing measurements focused on the removal of close galaxy pairs
(in 3D: \citealt{2002A&A...396..411K,2003MNRAS.339..711H}).  However, intrinsic
alignments can coherently anti-align galaxies that are well-separated along the line-of-sight.
\cite{2004PhRvD..70f3526H} highlighted the importance of this effect, and subsequent
observational work confirmed that it is the dominant impact of intrinsic alignments on weak
lensing measurements; but it cannot be eliminated by removing galaxy pairs at the same redshift from
the lensing sample.  Based on recent observational constraints \citep[e.g.,][]{2016MNRAS.457.2301S}, intrinsic alignments will be an important systematic
that surveys like LSST must mitigate (Figure~\ref{fig6}).  Efforts to remove this systematic typically include joint
modeling or self-calibration using joint analysis of galaxy-galaxy, galaxy-shear, and shear-shear
correlations \citep{2010A&A...523A...1J,2017arXiv170701072Y}. These approaches rely on the fact that
the various contributing terms have different redshift dependencies, spending some of the
statistical constraining power of the data to marginalize over the intrinsic alignments terms.

Current work on intrinsic alignments includes attempts at better observational constraints
(requiring redshift estimates and shape measurements), model building
\citep[e.g.,][]{2015JCAP...08..015B,2017arXiv170809247B}, and tests of mitigation methods.  
Of particular value would be large-area spectroscopic samples that would enable better priors to be
placed on the parameters of intrinsic alignment models at redshift $z\gtrsim 0.7$ and at typical
galaxy luminosities.  While the types of redshift samples that are often proposed as a solution to
the problem of calibrating the ensemble redshift distributions may be helpful in principle, it
depends on the survey layout (multiple small pencil-beam surveys can be used for redshift
calibration, while large-scale intrinsic alignments studies require larger area).
%

\subsection{Baryonic effects}


The impact of the physics of galaxy formation on weak lensing observables has been a subject of
study for more than a decade.  Unfortunately, thorough investigation of this topic requires
hydrodynamic simulations that have realistically complicated models of galaxy formation (without the
over-cooling problem), high enough
resolution to ensure their convergence for typical galaxy masses, and large enough volume to study
the impact on the matter power spectrum on cosmological distance scales.  This combination of
scenarios has only recently become possible, in families of very expensive high-resolution
simulations with box lengths of order 100~Mpc, including the EAGLE simulations
\citep{2015MNRAS.446..521S}, Illustris \citep{2014Natur.509..177V}, and MassiveBlack-II
\citep{2015MNRAS.450.1349K}.

One approach to account for the impact of baryons on the matter power spectrum is to include them
in a perturbation theory-based model for the power spectrum, with baryonic physics producing
higher-order terms that can be marginalized over \citep{2014MNRAS.445.3382M}.
There is also a halo model approach, which has nuisance parameters describing the change
in internal structure of dark matter halos, specifically their concentration, due to baryonic
physics 
\citep{2011MNRAS.417.2020S,2013PhRvD..87d3509Z}.  The extension of a halo model to quite small scales is simpler in
practice than the extension of the perturbation theory-based model mentioned above, which requires the inclusion of
many additional terms.  \citet{2013PhRvD..87d3509Z} found that for future lensing surveys, additional
mitigation may be needed, possibly reflecting the fact that a change in halo concentration is not
the only impact of baryonic physics. A halo model approach with changes in halo
concentrations  and a mass-dependent `halo bloating' parameter \citep{2015MNRAS.454.1958M}
has been quite successful in describing the matter power spectrum to small scales; the
parameters of that model were calibrated to maximize the fidelity of reproduction of the matter
power spectrum, rather than to accurately describe dark matter halo profiles.  This
approach was adopted by the KiDS survey \citep{2017MNRAS.465.1454H} to model shear-shear correlation
functions.

Rather than adopting a physically-motivated approach, \citet{2015MNRAS.454.2451E} used an empirical
PCA approach.  Using a set of cosmological hydrodynamic simulations to construct PCA
components that describe the impact of baryonic physics on the dark matter power spectrum, they
showed that excluding the first four PCA components is sufficient to mitigate the impact of baryonic
physics on a shear-shear correlation function measurement, even going to relatively small scales
($\ell\sim 5000$).

While current results seem to indicate that the impact of baryonic physics on shear-shear
correlations is under control for current surveys, and the methods that already exist are promising
for future (Stage-IV) surveys, there remain some questions in the field.  In particular, for the
tomographic shear-shear, shear-galaxy, galaxy-galaxy joint analysis that is likely the fiducial weak lensing
analysis for upcoming surveys (due to how it enables marginalization over shear-only systematics),
the production of theoretical models that self-consistently include the impact of baryonic physics
on all three correlation functions is less well-developed.  The halo model and extensions
of the PCA approach are promising avenues for investigation.  The scales on which these approaches
work for the full joint analysis is likely to determine the maximum usable
$\ell$ or minimum usable $\theta$.  Additional questions for investigation include the interaction
between marginalization over baryonic effects and other systematics.  For example, if intrinsic
alignment models are constructed separately for red and blue galaxies, such that theoretical models
separately predict the signal for the two populations and then take the appropriate weighted
averages, then does the baryonic physics model for galaxy-shear and galaxy-galaxy terms also need to
differ for the two populations?  This and other interactions (e.g., photometric redshift errors and
their uncertainties also
depend on the galaxy type) remain to be explored.

\subsection{Covariances}\label{subsec:covariances}


The process of inferring cosmological parameters given a set of measurements typically involves
knowing the covariance matrix of those measurements, under the assumption that the likelihood
function of the observable quantities is a Gaussian.  
Computing the covariance matrix for weak lensing measurements is a task for which multiple
approaches exist in the literature\footnote{Some of these approaches were developed for
shear-shear correlations, and the extension to galaxy-shear and galaxy-galaxy requires additional
work.}, and additional development will be needed for upcoming surveys.
In principle, future surveys may have data vectors with of order 1000 points, considering some
number of tomographic bins, bins in angular scale or wavenumber, and the three different types of
correlations to be measured.  Several studies have argued that the
number of simulation realizations of upcoming surveys needed to estimate the covariance matrix with
sufficient accuracy through brute force methods is prohibitively large
\citep{2013PhRvD..88f3537D,2013MNRAS.432.1928T}, though see recent work by
\citet{2017MNRAS.464.4658S} that argues those were significantly overestimated in the case that one
can parameterize the covariance in some compact way and use the simulations to constrain that
parameterization (see discussion below). 

The covariance matrix in general has shot noise terms and cosmic variance terms, including
contributions from connected four-point functions and supersample covariance
\citep{2014PhRvD..89h3519L,2014MNRAS.445.3382M}.  See \citet{2017MNRAS.471.3827S} for a recent
derivation of the generic covariance expression for two-point correlations of either densities and
overdensities, and quantities such as shear.  Because some of these terms are cosmology-dependent,
in principle, the covariance matrix itself should be re-estimated at each step of a
likelihood analysis to constrain cosmology \citep{2009A&A...502..721E}.

Numerical estimation of the covariance matrix using theoretical expressions is a
natural way to incorporate the cosmology-dependence of the covariance.  However, ensuring the
numerical stability of all terms in the 
covariance matrix estimation can be quite expensive
\citep[e.g.,][]{2017arXiv170609359K}.  Hence, building an emulation tool that would enable fast
estimation of these covariances would be highly valuable.  Most lensing analyses to date have not
incorporated a full cosmology-dependent covariance, with the notable exception of
\citet{2013ApJ...765...74J} for non-tomographic shear-shear correlations only.  In the KiDS analysis
that included shear-shear, shear-galaxy, and galaxy-galaxy correlations \citep{2017arXiv170605004V},
the cosmology-dependence of the covariance was partially accounted for through an iterative
procedure.  While they did not vary the covariance at each step of their MCMC, they did use the
best-fitting cosmology from their first MCMC to regenerate the covariance and then rerun the fitting
procedure.  

Another approach that has seen popularity with past surveys is direct empirical estimation of the
covariances, such as using the jackknife or bootstrap method, with the subsamples consisting of
large contiguous regions within the survey \citep[e.g.,][]{2013MNRAS.432.1544M}.  This approach has
been rigorously compared with both numerical estimates of covariances and with realistically complex
mock catalogs for galaxy-shear and galaxy-galaxy correlations
\citep{2017MNRAS.470.3476S,2017MNRAS.471.3827S}, and has been found to be quite accurate for scales
up to the size of the jackknife regions.  A natural tension for this method is that the need for the
number of regions to significantly exceed the number of data points motivates the use of many
smaller regions, but use of smaller regions reduces the range of scales on which the jackknife can
be accurately used, and causes a violation of the assumption of region independence.
However, if a given survey configuration allows the jackknife method to be used, it can be useful in
avoiding the need for many realizations of mock catalogs.  
The covariance matrix estimated in this way will be noisy, and since the inverse
covariance used for likelihood analysis is then biased, the sizes of cosmological parameter
constraints must be corrected 
\citep[e.g.,][]{2004MNRAS.353..529H,2007A&A...464..399H} .

In principle, using many simulation realizations of the survey provides a way to estimate 
covariance matrices.  Similarly to the above empirical methods, each element of the covariance
matrix must be independently constrained, and hence longer data vectors pose a greater challenge.
For expected data vectors in the surveys of the 2020s, the number of
realizations needed to do this as a function of cosmology is likely prohibitive, even assuming the
expected increase in computing power available in the 2020s.  The noise resulting from the limited
number of realizations compared to the number of data points \citep{2016MNRAS.458.4462B}
must also be taken into account, e.g., using the methods mentioned above for empirical covariances.
However, it is more likely that a hybrid approach will be used, adopting some method for modeling
the covariance and then constraining its (much smaller number of) parameters using the simulations.
For example, methods for modeling the precision matrix (inverse covariance matrix;
\citealt{2016MNRAS.460.1567P,2017arXiv170307786F}) or perturbation theory approaches to the
covariance \citep{2017MNRAS.466..780M} may be useful, in addition to simulation-calibrated versions
of the numerical models mentioned above.  Techniques for fast mode resampling may also be useful in
reducing the number of simulation realizations needed \citep{2011ApJ...737...11S}.
Finally, data compression methods introduced in Sec.~\ref{subsec:estimators} are also relevant
here. 
In the coming years, it will be important that a way forward that works for the full tomographic
shear-shear, galaxy-shear, and galaxy-galaxy analysis be validated and implemented; see
\citet{2015MNRAS.451.1418M} for a demonstration of some issues that arise when combining
galaxy-shear and galaxy-galaxy correlations and estimating covariances.  Ideally, a fast
emulator \citep[e.g., building on ideas from][]{2013JCAP...11..009M} would be constructed, to avoid 
cosmology-dependent covariance matrix estimation being a primary limiting step in the final
likelihood analysis.

\subsection{Inference}\label{subsec:inference}


Cosmological inference from weak lensing with past and current datasets has typically involved the
assumption that the likelihood function of the observables in Gaussian, and use of some form of
Markov Chain Monte Carlo \citep{2002PhRvD..66j3511L,2009MNRAS.398.1601F,2013PASP..125..306F} to
sample parameter space and identify best-fitting parameters and confidence intervals.
 It may also be
important for future 
surveys to account for non-Gaussianity in the likelihood
\citep[e.g.,][]{2009ApJ...701..945S,2011ApJ...734...76S}.

One of the key challenges facing future surveys in the cosmological parameter inference step is the
high dimensionality of the problem.  The multiple tomographic bins and three correlations to
jointly model produce of order $1000$ data points. The
number of systematics that must be marginalized over will be of order $100$, in addition to 
of order $10$ cosmological parameters.  While alternative inference methods have been proposed
\citep[e.g.,][]{2013ApJ...779...15J,2015A&A...583A..70L,2016MNRAS.455.4452A}, substantially more research must be done
to ascertain the feasibility of adopting them for future cosmological lensing surveys.  Some of these methods come with
the substantial benefit that one can avoid the Gaussian likelihood assumption and 
covariance matrix estimation, typically at the cost of requiring fairly realistic forward simulation
techniques for the observable quantities.

One issue that has received significant attention in the weak lensing community is confirmation bias
\citep[e.g.,][]{2011arXiv1112.3108C}, the solution for which is to carry out a blinded analysis
until null tests are passed and decisions have been made about what range of scales and models
to use.  All three ongoing weak lensing surveys have adopted a blind cosmological analysis strategy
 \citep{2017arXiv170801530D,2017MNRAS.465.1454H,CatalogPaper:inprep}.  The
details depend on the survey and the analysis being carried out, and most
surveys adopt a combination of the following: (a) applying a randomly selected
calibration factor to the shears, (b) having multiple catalogs with randomly selected calibrations
and only one person able to reveal which of those catalogs is the true one (zero additional
calibration factor), (c) applying random calibration factors to the measured two-point correlations,
(d) avoiding plotting the data against predictions from any cosmological model, (e) looking at MCMC
results only after subtracting off the best-fitting cosmological parameters, i.e.,
$\Delta\Omega_m$ rather than $\Omega_m$ itself.  Some of these strategies  may only work with
current datasets (where cosmological parameter changes look close enough like calibration offsets),
but not 
future data (where changes in shape of the two-point correlations with cosmology will be evident due
to the smaller errors).  The common adoption of blind analysis methods is a positive step forward
for the field of weak lensing, and current surveys should be quite informative as to which methods
are likely to work for future surveys.

\section{DETECTING AND MODELING OBSERVATIONAL SYSTEMATICS}\label{sec:obssys}


 In this section, I discuss the classes of systematic
errors, and the tests that can help reveal them using the data itself. 
 For survey papers that use many 
systematics tests to reveal observational systematics, including `null
tests' (which should be zero in the absence of systematics), see 
\citet{2017MNRAS.465.1454H,CatalogPaper:inprep,2017arXiv170801533Z}.
Here I will not discuss null tests for PSF modeling errors, which were thoroughly
discussed in Sec.~\ref{subsec:psfmodeling} of this review.
Before considering specific null tests, it is worth noting a general rule that null tests are often
most informative when carried out after binning samples based on  any independent quantity that could be
related to a potential systematic. 

As mentioned in Sec.~\ref{subsec:theory}, cosmological B modes are expected to be quite small,
and hence a detection of non-zero B-mode power is typically interpreted as arising from
systematics.  Unfortunately, B modes can have many origins, including PSF modeling
errors, PSF correction errors, astrometric errors, and intrinsic alignments.  Uncovering which of
these is responsible can be difficult, and
the correct mitigation scheme to use depends on the origin of the effect
\citep{2017MNRAS.465.1454H}.  Also, many systematics do not generate B modes, and hence a lack of
B modes does not guarantee a systematics-free measurement.

Another common diagnostic for additive biases, the star-galaxy correlation function, can be nonzero for a number of
reasons.  This test involves correlating the shapes of stars with the PSF-corrected galaxy shear
estimates.  Thus, both PSF modeling errors and insufficient PSF correction of galaxy shapes can contribute.  This test has
been used in different ways in previous lensing measurements.  The zero-lag star-galaxy correlation
can be estimated using the PSF model shape for the ``star'' shape, and averaged within small regions
\citep{2012MNRAS.427..146H,2017MNRAS.465.1454H}.  Assigning an uncertainty on this quantity
generally requires using mock catalogs that have a realistic level of cosmic variance and PSF model
variation across the fields.  This test was used in CFHTLenS to eliminate outlier fields that
(for undetermined reasons) were too systematics-dominated to use for science.  
An alternative approach \citep{CatalogPaper:inprep} is to measure the full
star-galaxy shape correlation as a function of separation, averaged over the entire survey.  In
principle, this should be the sum of terms from PSF modeling errors (related to the $\rho$
statistics) and from uncorrected PSF anisotropy.  It provides a template for marginalization over
additive errors due to these systematics; however, leakage across the star-galaxy boundary can
result in this correlation including cosmic shear as well.

Correlations with systematics maps is another method that can enable the detection of observational
systematics \citep{2017arXiv170801535C,Oguri:inprep}.  This method involves producing lensing mass maps from
the shear catalog, and maps corresponding to the values of any quantity that may be considered as a
possible cause for weak lensing systematics (e.g., stellar density, PSF FWHM, PSF shape).  The
cross-correlation between the lensing and systematics maps should be zero in the absence of systematics.  Map-level
correlations can be a more compact way to detect certain systematics, rather than re-computing all
2-point correlation null tests after dividing the sample into bins in seeing and other quantities
\citep[as was done in, e.g.,][]{2016PhRvD..94b2002B}.

Calculating average shears with respect to arbitrary locations that should not generate lensing shear is
another common null test.  For example, the average shapes of galaxies with respect to the CCD
coordinate system or the positions of stars should be zero (modulo noise and the contamination of the star sample with galaxies).  The caveat in
the parenthesis highlights another important point, which is that the origins of deviations from
zero for null tests should be carefully considered.
Sometimes the source of the signal observed is completely different from what was originally intended.

There are few null tests that are meant to specifically identify residual detector
effects.  A recent example is the computation of PSF model size residuals as a
function of stellar magnitude \citep{CatalogPaper:inprep,2017arXiv170801533Z}.  Computing the mean
shear in CCD coordinates for galaxies binned based on their CCD row/column can also be useful for
identifying detector systematics \citep[e.g.,][]{2014MNRAS.440.1296H,2017arXiv170801533Z}. The
development of more tests that can 
identify failure to correct for detector effects or chromatic PSF effects would be
useful for the next generation of surveys, which require a greater level of control over those
effects.

One useful tool to detect systematics due to nearby galaxies and/or due to failures in the analysis
pipeline is to inject fake galaxies into real data and rerun the analysis pipeline
\citep{2016MNRAS.457..786S,2017arXiv170501599H}.  Comparison of the measured properties of the fake
galaxies with the input ones can help diagnose problems with many steps of the
analysis pipeline (detection, deblending, photometry, shear estimation).  The
impact of the injected galaxies on the real ones can also be measured; while we do not know ground
truth for the real galaxies, the difference between the originally-measured properties and those measured after injection
of the fake objects can be revealing \citep{2017arXiv170801534S}.

Unfortunately, there is no observational test to identify failures in the absolute multiplicative
calibration of the ensemble shear signal, which is why the problem described in
Section~\ref{subsec:shearest} has attracted so much attention.  However, comparing subsamples of
galaxies can reveal relative calibration biases between subsamples, modulo selection bias
(Section~\ref{subsec:selbias}), which  makes the division into subsamples a
potentially problematic null test. In other words, for this to be a useful test, the standard
sources of bias such as noise bias, model bias, and selection bias must be separately calibrated out
for the subsamples in order to use this as a test for unrecognized/unknown systematics. 
One important aspect of shear comparisons (whether between subsamples within a given survey, or between the same sample of galaxies measured
in two surveys) is that they should always happen at the level of ensemble shears, not per-galaxy shapes,
for the reasons explained in Section~\ref{subsec:shearest}. See \citet{2017arXiv170704105A} for one example of a recent
shear comparison, with methodology that should be applicable elsewhere.  This comparison can be done at the level
of ensemble shear estimates for matched samples with a common set of photo-$z$'s, to identify 
shear-related calibration offsets, or at a higher level that 
includes both photo-$z$ and shear-related calibration offsets.

To identify and remove additive systematics due to physical effects associated with specific
exposures or surveys (e.g., atmospheric PSFs, or a detector effect), one possible way
forward is to cross-correlate shear maps from different surveys or subsets of exposures
within a single survey.    In principle, this test could be extended not just to consistency
tests (i.e., split the LSST exposures into two sets, and do a separate analysis in each one), but to
detect and exclude data with potentially unknown systematic errors (i.e., through a
jackknife process that involves sequentially excluding small portions of data and testing their
statistical consistency with the rest).

The above tests (and the $\rho$ statistics described earlier in this review) and analysis of survey
simulations can be used to identify the presence of systematics that can contaminate
cosmological weak lensing analysis.  They will also provide templates
for residual systematics that can be marginalized over when constraining cosmological parameters.
It is important that these template not only be scale-dependent, but also galaxy property-dependent
and/or redshift-dependent, since most
shear systematics will depend on the galaxy properties at some level, and hence on the redshift.
Indeed template marginalization is a popular method for removing theoretical systematics, but unlike
for observational systematics, there are fewer null tests that can be done for theoretical
systematics (with typical tests being eliminating data in regions where the systematics should be
worse, and testing for consistency of results).

\section{SUMMARY}\label{sec:summary}


The high-level goal of this review is to answer the following question: ``What does the weak lensing
community need to do in order to get to the point where surprising claims that are made
about dark energy with LSST, Euclid, or WFIRST will be believed?'' 

The fact that LSST, Euclid, and WFIRST are designed in ways that result in different
dominant systematics is an important aspect of the landscape of the 2020s.  For example, Euclid's
broad RIZ filter means that it is far more susceptible to chromatic effects
(Section~\ref{subsec:psfmodeling}) than LSST or WFIRST.  The WFIRST survey design is more
conservative than Euclid in terms of the number of samples at each position, making it less likely
to suffer from undersampling (due to, e.g., cosmic rays that result in an exposure being excluded).
WFIRST's NIR detectors will have different pixel-level systematics than the Euclid or LSST CCDs, and
greater calibration challenges due to the CMOS architecture.
LSST will suffer from blending far more than the two space-based surveys.  The fact that Euclid is
shallower than WFIRST or LSST means that it can more easily gather representative spectroscopic
samples for photometric redshift training and calibration.  However, the fact that WFIRST will be
completely within the LSST footprint (and will use it for photometric redshift determination)
results in greater survey homogeneity than Euclid, which will rely on 
several ground-based datasets for photometric redshifts.  Relying on the various survey 
cross-comparisons plus the fact that they suffer from systematics at different levels will be highly
scientifically valuable, and the combination of the surveys has the potential to be even more
powerful than one would expect by naively combining statistical errors
(\citealt{2015arXiv150107897J}, Rhodes et~al.\ {\em in prep.}).

Below are a number of key take-aways synthesizing the material in the sections above:
\begin{enumerate}
\item There are low-level issues such as detector systematics, chromatic effects,
  astrometry, and survey geometry representation for which work is clearly needed to
  get where we need to be for surveys in the 2020s, but there are promising avenues for investigation.
\item An area where genuinely new ideas are needed is blending systematics, both in how to
  quantify and mitigate the impact of low-level blends on shear and photo-$z$, and the impact of
  unrecognized blends.  The field has only recently started to confront this issue, and more work is needed.
\item Several issues fall into the category of ``promising ideas exist but more exploration is
  needed to determine which will work and how exactly to use them at the level of precision needed
  for future surveys'': calibration of $N(z)$ for photometric redshift samples, shear calibration,
  optimal image combination, PSF modeling, mitigation of theory systematics, and covariance matrix estimation.
  Serious work must be done by the community, but all of these issues are more advanced than
  blending systematics.  The calibration of $N(z)$ for photometric
  redshift samples has gotten less attention than shear calibration until recently, and therefore
  there is some catching up to do in this area.  Indeed, the weak lensing
  community's unfortunate habit of outsourcing photo-$z$ production and calibration without
  considering the 
  cross-talk between shear-related selection effects and photo-$z$'s must 
  end: we must interface with the photo-$z$ community at an earlier phase of the analysis.
\item Decisions to be made about image combination must factor in the connection between image
  combination, PSF modeling, shear estimation, and deblending.
\item Choice of data compression methods will have an impact on the best way to handle covariance
  matrix estimation and cosmological parameter inference.
\item The field views shear estimation quite differently from how it did from the mid-1990s until
  around 2012: it is now
  well-understood that estimation of per-galaxy shapes will not result in an unbiased
  estimate of the ensemble shear, so the focus is on either calibration
  strategies or methods of inferring shear without per-galaxy shapes.  Several highly promising
  options currently exist.
\item Regarding the overall cosmological inference problem, more work is needed on blinding
  strategies for weak lensing analysis by upcoming surveys.  In
  addition, there is still room to draw the field away from the standard method of likelihood
  analysis (see alternatives discussed in Sec.~\ref{subsec:inference}), but it will take substantial
  development for those methods to be viable.
\item Having (at least) two methods with different
  assumptions for any complex analysis step is highly valuable.  This
  was highlighted in the DES year 1 cosmology analysis \citep{2017arXiv170801530D}.  Even
  having two independent pipelines that share assumptions can be useful for
  identifying bugs, hidden assumptions, and numerical issues.  Pipeline redundancy will likely remain an
  important element of cosmology analysis in the 2020s, and hence it is really valuable that for
  most of the key issues discussed here (e.g., $N(z)$ and shear calibration) there are
  multiple viable approaches.
\item Null tests are valuable, but it is important to understand what really is a null test, and
  which ``null tests'' could be defeated by faulty assumptions.
\item In many of the above sections on theoretical systematics, papers that are referenced show
  methods for marginalizing over that systematic. Most of those papers considered individual
  systematics in isolation.  The full problem with all of these theoretical and
  observational systematics is likely more complex, with degeneracies between some systematics. 
  It will be important for the field to confront the multiple-systematics-mitigation problem sooner
  rather than later, in order to identify obstacles early on and develop strategies to overcome them.
\end{enumerate}

The issue of multiple independent approaches raises the question of how independent the different
surveys should try to be.  For example, if they all rely on the representativeness of a given
spectroscopic training sample for photo-$z$, systematic uncertainties could become correlated across
the surveys. Cross-survey comparison, when carried out properly, can be a powerful tool for
identifying inconsistencies that may be due to systematic errors.  But it is important to consider
the ingredients of the analysis (e.g., commonality in algorithms for shear calibration or photo-$z$
calibration, same vs.\ different implementations of common algorithms) before deciding what parts
truly are independent.  In that sense, comparison against CMB lensing is ``safer'' as a
cross-check, because it is difficult to correlate systematics between galaxy and CMB lensing.
It is clear, however, that a broad set of internal cross-checks (Section~\ref{sec:obssys}) and
external ones will be necessary for the surveys of the 2020s to produce credible weak lensing
cosmology results.  This work should begin before the 2020s: existing surveys -- KiDS, HSC, and DES
-- will play a crucial role in this path towards believable 
precision cosmology with the surveys of the 2020s.  The community must demonstrate an ability to
self-consistently constrain cosmology with these datasets.

While Figure~\ref{fig:forecast} (left panel) provided an initial motivation for why weak lensing is
so valuable as a cosmological probe, the sections above may raise the question of why try to do it
at all given the complexity of the problems involved?  The 
community has made tremendous strides in how to address the key problems facing the field, and most
outstanding issues now have multiple paths to a resolution.  To distinguish between general dark
energy and modified gravity models as the cause of the observed accelerated expansion rate of the
Universe, we generally require a probe of the distance-redshift relation (e.g., baryon acoustic
oscillations, supernovae, time-delay strong lenses) and structure growth (weak lensing, galaxy
clustering, redshift-space distortions, galaxy cluster counts).  While all come with challenges,
weak lensing is currently the most promising of the ``structure growth'' probes.  Use of galaxy
clustering or redshift-space distortions alone requires highly precise determination of the galaxy
bias or marginalization over its value (which weakens constraints).  In contrast, weak lensing
allows direct determination of the galaxy bias from the shear-shear, galaxy-shear, galaxy-galaxy
correlations.  Competitive galaxy cluster measurements of cosmology require weak lensing
measurements with special care for systematics and theoretical uncertainties 
arising in crowded cluster regions. All probes of structure growth besides shear-shear correlations suffer
worse from baryonic effects, since weak lensing signals are dominated by collisionless matter in the
translinear regime. In short, these factors plus the tremendous development in the field of weak
lensing in the past decade lead to the conclusion that weak lensing provides the cosmology
community's best hope for competitive and believable constraints on cosmic structure growth, and
hence on dark energy.


\section*{ACKNOWLEDGMENTS}


I would like to thank Mike Jarvis, Joe Zuntz, and Daniel Gruen for providing helpful feedback on the
structure of this review based on early outlines.  I also thank Gary Bernstein, Jim Bosch, Scott
Dodelson, Mike Jarvis, Benjamin Joachimi, Tod Lauer, Peter Melchior, Josh Meyers, Jeff Newman, Sam Schmidt, Michael
Schneider, Chaz Shapiro, Erin Sheldon, and Michael Troxel for
reading portions of this review and providing thoughtful feedback on relatively short notice.

%

\bibliographystyle{ar-style2}

\bibliography{wl,papers,ia}

\end{document}